\documentclass[12pt,preprint]{aastex}

\usepackage{natbib}

\slugcomment{Last edited: \today}

\shorttitle{H$\alpha$ imaging of UCM galaxies}
\shortauthors{P\'erez-Gonz\'alez et al.}

\begin{document}

\title{SPATIAL ANALYSIS OF THE H$\alpha$ EMISSION IN THE LOCAL STAR-FORMING UCM GALAXIES}

\author{Pablo G. P\'erez-Gonz\'alez\altaffilmark{1}, Jaime Zamorano and Jes\'us Gallego}
\affil{Departamento de Astrof\'{\i}sica, Facultad de F\'{\i}sicas, Universidad Complutense, E-28040 Madrid, Spain}

\author{Alfonso Arag\'on-Salamanca}
\affil{School of Physics and Astronomy, University of Nottingham, NG7 2RD, UK}

\and 

\author{Armando Gil de Paz}
\affil{The Observatories of the Carnegie Institution of Washington, 813 Santa Barbara St., Pasadena, CA 91101, USA}

\altaffiltext{1}{pag@astrax.fis.ucm.es}

\begin{abstract} We present a photometric study of the H$\alpha$ emission in
the {\it Universidad Complutense de Madrid} (UCM) Survey
galaxies. This work complements our previously-published spectroscopic
data. We study the location of the star-forming knots, their
intensity, concentration, and the relationship of these properties
with those of the host galaxy. We also estimate that the amount of
H$\alpha$ emission that arises from the diffuse ionized gas is about
15--30\% of the total H$\alpha$ flux for a typical UCM galaxy. This
percentage seems to be independent of the Hubble type. Conversely, we
found that an `average' UCM galaxy harbours a star formation event
with 30\% of its H$\alpha$ luminosity arising from a nuclear
component.  The implications of these results for higher-redshift
studies are discussed, including the effects of galaxy size and the
depth of the observations.  A correlation between the SFR and the
Balmer decrement is observed, but such correlation breaks down for
large values of the extinction. Finally, we recalculate the H$\alpha$
luminosity function and star formation rate density of the local
Universe using the new imaging data. Our results point out that, on
average, spectroscopic observations detected about one third of the
total emission-line flux of a typical UCM galaxy.  The new values
obtained for the H$\alpha$ luminosity density and the star formation
rate density of the local Universe are
$10^{39.3\pm0.2}$~erg\,s$^{-1}$\,Mpc$^{-3}$, and
$\rho_{SFR}=0.016^{+0.007}_{-0.004}$\,$\mathcal{M}_\sun$\,
yr$^{-1}$\,Mpc$^{-3}$ ($H_{0}=50$\,km\,s$^{-1}$\,Mpc$^{-1}$,
$\Omega_{\mathrm M}=1.0$, $\Lambda=0$). The corresponding values for
the `concordance cosmology' ($H_{0}=70$\,km\,s$^{-1}$\,Mpc$^{-1}$,
$\Omega_{\mathrm M}=0.3$, $\Lambda=0.7$) are
$10^{39.5\pm0.2}$~erg\,s$^{-1}$\,Mpc$^{-3}$
$\rho_{SFR}=0.029^{+0.008}_{-0.005}$\,$\mathcal{M}_\sun$\,yr$^{-1}$\,Mpc$^{-3}$.

\end{abstract}

\keywords{surveys --- galaxies: starburst ---    galaxies: fundamental parameters --- 
galaxies: photometry ---  galaxies: stellar content}

\section{Introduction}

The characterization of the star formation in galaxies at any cosmic
epoch remains a topic of considerable interest
\citep[see, e.g.,][]{1999ApJ...519L..47Y,2000ApJ...544..641H,2001MNRAS.320..504S,
2001ApJ...549..745R,2002ApJ...570..492L}. Much effort has been devoted
to the study of star-forming galaxies at low, intermediate, and high
redshift in the last decade \citep[among
others,][]{1994ApJS...94..461S,
1995ApJ...455L...1G,1996ApJ...460L...1L,1996MNRAS.280..235E,
1998ApJ...504..622H,2001ApJ...550..593J,2001ApJ...554..981P,2001A&A...379..798P,
astro-ph/0111390}. These studies have revealed that the comoving star
formation rate (SFR) density of the Universe has increased by an order
of magnitude from the present time to $z\sim1-2$. The behavior beyond
redshift $z\sim2$ is still an open issue, with some results indicating
a declining SFR with redshift, while others suggest little or no
evolution.

If we want to understand how the star-formation process has evolved with time,
the statistical determination of the star formation history of the Universe as
a whole needs to be complemented with the  characterization of the
star-formation properties of the individual galaxies. This is not an easy task
for high-$z$ objects,  given their small angular size and faintness, but some
progress is being made. Complementary studies of local samples of star-forming
galaxies are also clearly needed for comparison purposes, and to provide
information on the end-point of the evolution. Thus,  detailed analysis of the
spatial distribution and intensity of the star formation in local  galaxies may
have important implications for studies at intermediate and high redshifts.  
Though evolution certainly plays an important role when  studying galaxies at
different redshifts, some properties may remain comparable and easily
measurable \citep{1998Ap&SS.263....1G,2000A&A...354L..21C}. Several authors
have found similarities between local and distant objects in size
\citep{2001hsa..conf...97P}, intensity of the starburst
\citep{1997AJ....114...54M}, relationship between the  newly-formed stars and
the population of more evolved stars \citep{2003MNRAS.338..525P},
stellar mass \citep{astro-ph/0204055K}, etc.  In addition, much work
is now focusing on the characterization of local counterparts of
distant objects (e.g., the Luminous Blue Compact Galaxies in
\citealt{1997ApJ...489..559G} and \citealt{2001AJ....122.1194P}, or the UV-bright 
starburst galaxies in \citealt{2000AJ....119...79C}).

\citet{1983ApJ...272...54K} carried out  pioneering work on the study of the
structural properties of the recent star formation in normal disk
galaxies.  This work made use of the H$\alpha$+[\ion{N}{2}] nebular
emission as a tracer of the population of young hot stars (responsible
for the heating of the gas).  Similar papers have gone deeper into the
issue, following a variety of approaches: (1) comparing the rate of
star formation with other properties such as the morphology, the gas
content or the older stellar population content
\citep{1994ApJ...435...22K,1998ApJ...498..541K}; (2) shedding light on  the
influence of the environment on star formation by studying samples of field
galaxies, compact groups, and clusters
\citep{1998AJ....115.1745G,1999ApJ...518...94I,1999AJ....118..730H}; (3) 
extending the analysis to galaxies with enhanced star formation
\citep{2000AJ....119...79C,2001JApA...22..155C}. These studies demonstrate 
that the use of H$\alpha$ photometry can be extremely useful in the
characterization of the youngest stellar population in star-forming galaxies,
and in complementing spectroscopic observations.

Another interesting topic outlined relatively recently is the amount of
H$\alpha$ emission arising from the {\it diffuse ionized gas} (DIG) of galaxies
\citep{1990IAUS..139..157R}. The DIG has been studied in nearby normal spiral
galaxies with high angular resolution \citep[see, e.g.,][]
{1994ApJ...426L..27L,2001ApJ...559..878H,2002A&A...386..801Z}. The
ionizing source of this interstellar medium component is still
uncertain: some authors identify it with young hot stars in HII
regions, while others introduce new possibilities such as supernova
remnants, turbulent mixing layers or even the extragalactic radiation
field \citep{1997ApJ...491..114W}. If the link between the DIG and the
HII regions is confirmed, models should account for the escape of a
non-negligible fraction of ionizing Lyman photons, which could reach
the intergalactic medium. This effect, if confirmed, would imply that
the factor used when converting from H$\alpha$ luminosity to star
formation rate could be underestimated \citep[see][and references
therein]{1998ARA&A..36..189K}.

In this paper we present H$\alpha$ imaging observations for a complete sample
of local star-forming objects: the {\it Universidad Complutense de Madrid}
(UCM) Survey galaxies \citep{1994ApJS...95..387Z,
1996ApJS..105..343Z,1999ApJS..122..415A}.  These galaxies present enhanced star
formation in comparison with normal quiescent spiral galaxies
\citep{2000MNRAS.316..357G,2003MNRAS.338..508P,2003MNRAS.338..525P}. A 
comparison of the UCM Survey galaxies with distant objects has been
carried out for different properties, including size, total luminosity
\citep{2000A&AS..141..409P}, mass, and intensity of the star formation episode
\citep{2003MNRAS.338..525P}. This sample has often been used as the local
benchmark for intermediate and high redshift studies \citep[see, for
example,][]{1998ApJ...498..106M}.

The UCM galaxies have been extensively observed in optical and near
infrared photometric passbands
\citep{1996A&AS..118....7V,2000MNRAS.316..357G,2000A&AS..141..409P,
2003MNRAS.338..508P}.  Optical long-slit spectroscopy
\citep{1996A&AS..120..323G} is also available for the entire
sample. The latter study allowed to obtain emission-line fluxes and
ratios, yielding results on star formation rates (via H$\alpha$
fluxes), extinctions (through the Balmer decrement), metallicities
(using oxygen abundances), etc. However, these observations suffer
from the effect of the finite aperture of the slit and the consequent
loss of spatial information. In this sense, spectroscopic observations
(commonly centered in the galaxy nucleus\footnote{In this paper we
will refer to the starbursts in the inner parts of spirals aside
nuclear activity as nuclear star formation.}) may have missed large
HII regions in the outer disk zones. In addition, when aperture
corrections are applied, a strongly emitting nucleus may cause the
integrated emission to be largely overestimated. The H$\alpha$ imaging
data presented in this paper complement the spectroscopic
observations, providing information about the location of the
star-forming knots, their intensity and relationship with the host
galaxy where the starburst is taking place. In addition, an estimate
of the DIG emission will be measured for a complete sample of
star-forming galaxies.

This paper is structured as follows. Section~\ref{sample} presents the
sample of galaxies and observations used in this work. Next,
Section~\ref{results} discusses on the results concerning the
comparison between imaging and spectroscopy, the concentration and
size properties, the DIG characterization, the implications for
studies at higher redshifts, and the corrections to the local
H$\alpha$ luminosity function and star formation rate density. Unless
otherwise indicated, throughout this paper we use a cosmology with
$\mathrm H_{0}=70$\,km\,s$^{-1}$\,Mpc$^{-1}$, $\Omega_{\mathrm M}=0.3$
and $\Lambda=0.7$.

\section{The sample}
\label{sample}

The {\it Universidad Complutense de Madrid} (UCM) Survey for
emission-line galaxies was carried out in the late 1980s and early
1990s with two main goals: (1) characterizing the star formation in
the local Universe, and (2) studying low metallicity objects
(\citealt[Lists I]{1994ApJS...95..387Z}; \citealt[List
II]{1996ApJS..105..343Z}; and \citealt[List
III]{1999ApJS..122..415A}). Extensive imaging and spectroscopic
follow-up observations are available for the vast majority of the
galaxies in the sample.

Long-slit spectroscopy for the UCM galaxies was presented in
\citet{1996A&AS..120..323G,1997ApJ...475..502G}. The mean redshift of the
complete sample of 191 emission-line galaxies is 0.026.  The sample is
dominated (57\%) by objects with low excitation, relatively high
metallicity, and often bright and dusty nuclear starbursts. We call
these disk-{\it like} galaxies.  A significant number of high
excitation, low metallicity galaxies are also present (32\%), and we
call them HII-{\it like} galaxies. These have blue star-forming knots
which may sometimes dominate the optical luminosity of the whole
galaxy, as in the case of {\it Blue Compact Galaxies} (BCDs). A number
of AGNs (8\%) was also detected. For the remaining 3\% it was not
possible to obtain a spectroscopic classification. The spectroscopic
data were also used to obtain the luminosity function and star
formation rate density of the local Universe
\citep{1995ApJ...455L...1G,2002ApJ...570L...1G}.

Imaging and photometric studies in the Gunn-$r$
\citep{1996A&AS..118....7V,1996A&AS..120..385V} and Johnson-$B$ bands 
\citep{2000A&AS..141..409P,2001A&A...365..370P} have also been carried out. The
average magnitudes of the sample are $m_B=16.1\pm1.1$ ($M_B=-19.2$)
and $m_r=15.5\pm1.0$ ($M_r=-19.8$), with a mean effective radius of
2.8 kpc (in $B$). Up to 65\% of the sample has been classified as Sb
or later. Near infrared data, stellar masses and the properties of the
youngest stellar population were presented in
\citet{2000MNRAS.316..357G} and
\citet{2003MNRAS.338..508P}.

This paper presents the narrow-band ($NB$) imaging in the redshifted
H$\alpha$ passband for a subsample of 79 galaxies from the UCM
Survey. The subset represents $\sim40$\% of the 191 galaxies within
the Lists I and II of the survey. These galaxies were selected to
ensure that all the spectroscopic types described in
\citet{1996A&AS..120..323G} were represented in the same proportion as
in the parent sample. The representativeness of this subsample in
H$\alpha$ equivalent width ---$EW(\mathrm H\alpha)$--- and luminosity
has also been tested. A Kolmogorov-Smirnov test using $EW(\mathrm
H\alpha)$ (a good tracer of the spectroscopic type) for the subsample
and the entire sample estimates the probability of the former being
representative of the latter in 82\%. The same test using the
H$\alpha$ luminosity yielded a probability of 98\%. Table~\ref{table1}
presents the sample along with the main results from this H$\alpha$
imaging study. For specific details on individual objects, the reader
may refer to \citet{2003MNRAS.338..508P} and references therein.

The 79 galaxies used in the present paper were observed in three
different campaigns and two service nights at the Nordic Optical
Telescope\footnote{The Nordic Optical Telescope is operated on the
island of La Palma jointly by Denmark, Finland, Iceland, Norway, and
Sweden, in the Spanish Observatorio del Roque de los Muchachos of the
Instituto de Astrof\'{\i}sica de Canarias.} and two runs at the CAHA
2.2 metre telescope\footnote{German-Spanish Astronomical Centre, Calar
Alto, operated by the Max-Planck-Institute for Astronomy, Heidelberg,
jointly with the Spanish National Commission for Astronomy.}. General
information about the campaigns is provided in Table \ref{observ}.

\placetable{observ}
\clearpage
\begin{deluxetable}{lcccccl}
\tablecaption{Log for the H$\alpha$ observations.\label{observ}}
\tablehead{
\colhead{Telesc./Observ.} & &\colhead{Date}& & \colhead{Instrument} & \colhead{Scale}&\colhead{Conditions}\\
 \colhead{(1)} & &\colhead{(2)}& & \colhead{(3)} &
 \colhead{(4)}&\colhead{(5)}\\}
\startdata
NOT (service)   &  Jul &13     & 1999 & HiRAC  &    0.110   & photometric\\
   2.2m CAHA    &  Dec &27--28 & 1999 & CAFOS  &    0.313   & cloudy\\
      NOT       &  Sep &22--25 & 2000 & ALFOSC\tablenotemark{a}  &    0.189   & 1 photometric night\\
NOT (service)   &  Sep &16     & 2000 & ALFOSC  &    0.189   & photometric\\
   2.2m CAHA    &  May &10--13 & 2001 & CAFOS  &    0.313   & non-photometric\\
      NOT       &  Jun &22--26 & 2001 & ALFOSC  &    0.189   & photometric\\
      NOT       &  Apr &19--21 & 2002 & ALFOSC  &    0.189   & 2 photometric nights\\
\enddata  
\tablenotetext{a}{The data presented here have been taken using ALFOSC, which is owned by the Instituto de Astrof\'{\i}sica de Andaluc\'{\i}a (IAA) and operated at the Nordic Optical Telescope under agreement between IAA and the NBIfAFG of the Astronomical Observatory of Copenhagen.}
\tablecomments{Observing log for the H$\alpha$ observations of 
the UCM Survey galaxies.  Columns stand for: (1) Telescope name. (2)
Date of the observation. (3) Instrument used (all equipped with a
Loral 2k$\times$2k detector). (4) Scale of the CCD in
arcsec~pixel$^{-1}$. (5) Weather conditions.}
\end{deluxetable}
\clearpage

The observations included narrow-band imaging centered at the
wavelength of the redshifted H$\alpha$ emission-line of each galaxy,
and broad-band (Cousins-$R_\mathrm{C}$) imaging. The broad-band data
was used to subtract the continuum of the narrow band images. Typical
exposure times were 900 seconds in $R_\mathrm{C}$ and 2700s in the
$NB$, allowing us to reach a limiting surface brightness in H$\alpha$
(2$\sigma$) of
$\sim5\times10^{-17}$erg~s$^{-1}$~cm$^{-2}$~arcsec$^{-2}$. The total
exposure was split in several sub-exposures for cosmic ray and
artifact elimination. About $90$\% of the frames were taken in
subarcsecond seeing conditions.

Some of the data reduction and measurements were carried out using
several IRAF tasks\footnote{IRAF is distributed by the National
Optical Astronomy Observatories, which is operated by the Association
of Universities for Research in Astronomy, Inc. (AURA) under
cooperative agreement with the National Science Foundation.}, mainly
{\it ccdproc}, {\it imcombine} and {\it geomap}. Other procedures were
specially built for this work, including the fringing correction,
continuum subtraction, and flux measurement programs.  Standard steps
were followed in bias subtracting and flat-fielding. Some of the broad
band data presented a small level of fringing (1 count per 1000),
which was eliminated using an object-free pattern built as a
combination of all the frames taken during each night. Scaling of the
pattern was performed for each image. After reduction, the images were
aligned using the positions of $\sim10$ stars, and all the frames of
the same galaxy obtained with the same filter were combined.


Emission-line images were obtained by subtracting the continuum. This
was achieved through the comparison of at least 20 field stars per
frame whose emission-line fluxes were forced to be null in the final
frames. Based on the standard deviation of the ratios between the
fluxes of the stars in the $NB$ and the broad-band images, we estimate
a $\sim5$\% error in the subtraction of the continuum. The procedure
left a residual sky background in the H$\alpha$ images, which was
subsequently subtracted.

\citet{1992AJ....104..340L} stars were observed in order to  calibrate the
$R_\mathrm{C}$ frames. We included a $B-R_C$ colour term in the
calibration.  $B-r$ colours were taken from
\citet{2000A&AS..141..409P} and converted to $B-R_C$ using a mean
value of $r-R_C=0.36$ \citep{1995PASP..107..945F}. This colour is
expected to vary in less than $0.04$ magnitudes from one morphological
type to another. Standard spectrophotometric stars
\citep{1990AJ.....99.1621O,1992PASP..104..533H} were also observed to check the
values of the scale factor between the $R_\mathrm{C}$ and $NB$ filters. Images
taken during non-photometric nights were calibrated with short $R_\mathrm{C}$
exposures obtained in other campaigns during photometric conditions. 

In addition, 8 frames without photometric observations were
calibrated using the Gunn-$r$ data given in
\citet{1996A&AS..118....7V}. In order to do that, we first determined
the correlation between the $R_\mathrm{C}-r$ and $B-r$ colours in
$5\arcsec$ apertures with the data of the galaxies observed in
photometric conditions, obtaining
$R_C-r=(-0.238\pm0.038)+(0.011\pm0.047)\,(B-r)$. The unbiased standard
deviation of the correlation is $0.08^m$. This relationship was then
used to calculate the $R_C$ magnitudes and to calibrate the frames
obtained in non-photometric conditions. In order to estimate the
magnitude of the calibration uncertainties, 2--3 common objects were
observed in all the campaigns.

The transmission curves for all the $NB$ filters used (and also $R_C$) were
obtained from the observatories. Typical widths for the $NB$ filters were 50
\AA. This value is not small enough to completely avoid  contamination from
the [\ion{N}{2}]$\lambda6548$ and [\ion{N}{2}]$\lambda6584$ lines.
However, the transmissions at these wavelengths were often rather low
in comparison with that of the H$\alpha$ line (1/2 on average). Pure
H$\alpha$ fluxes were obtained taking into account these filter
transmissions for the three lines, the [\ion{N}{2}]/H$\alpha$ ratios
given by \citet{1996A&AS..120..323G} for each object, and the scale
factors between the $R_\mathrm{C}$ and $NB$ images obtained with field
stars. To estimate the extinction corrections, the spectroscopic
$\mathrm H\alpha/\mathrm H\beta$ ratios in
\citet{1996A&AS..120..323G} were used. These ratios were corrected for a
stellar absorption of 3\AA\, \citep{1998ApJS..116....1T,1999ApJS..125..489G}
and converted to $E(B-V)$ using the Galactic extinction curve of
\citet{1989ApJ...345..245C}.

\begin{figure}
\plotone{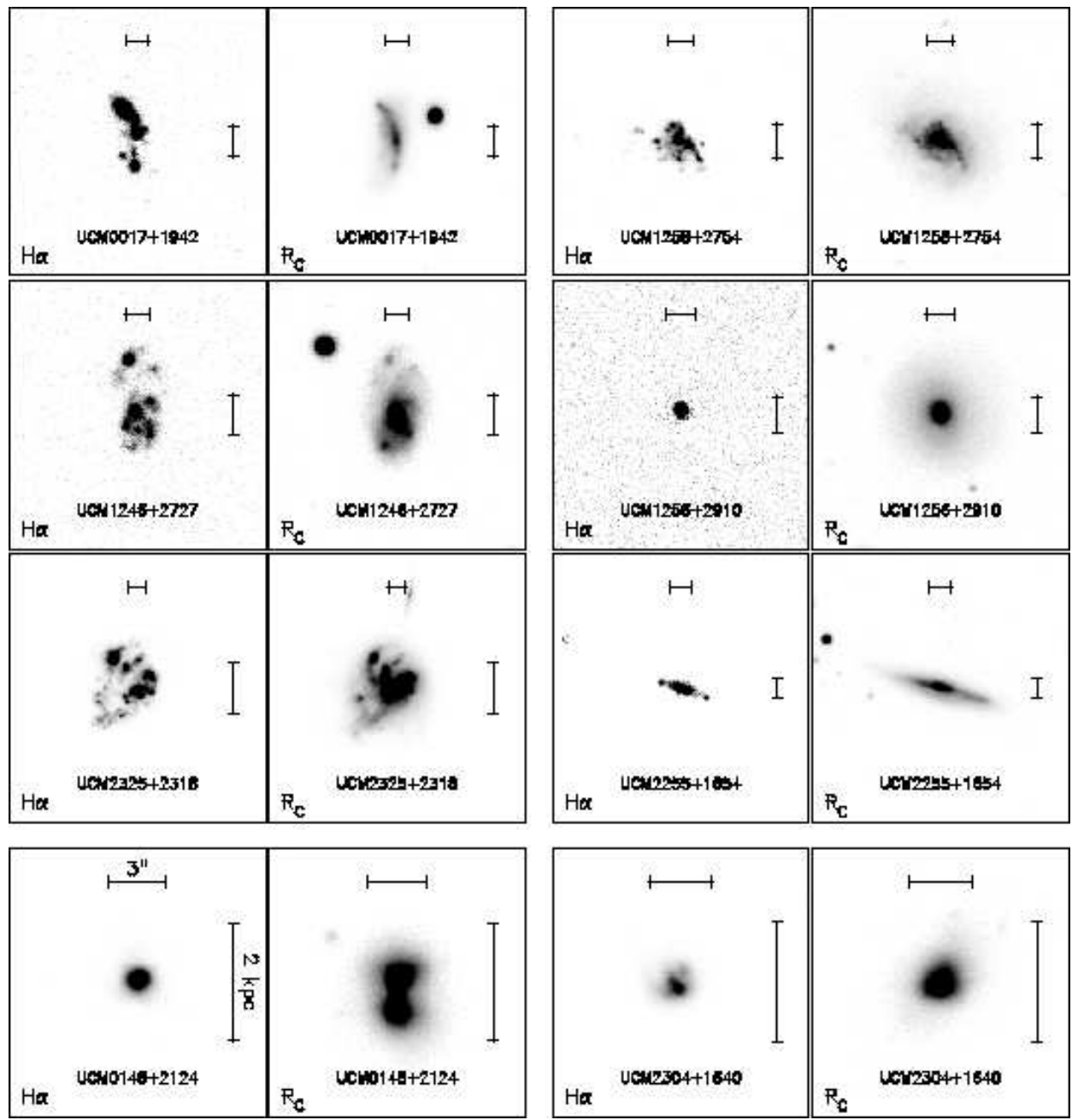}
\figcaption{\label{some}Gray-scale representation of some of the images  
of the UCM Survey galaxies. For each object, H$\alpha$ and
Cousins-$R_C$ images are shown. North is up and east is left. An
angular size of $3\arcsec$ is marked at the top of each image
(horizontal segment). A 2~kpc transversal distance is also depicted on
the right of each image (vertical segment). The 3 first rows plot
HII-{\it like} galaxies (on the left) and disk-{\it like} objects
(right). The 2 objects in the last row are BCDs. The images for the
entire sample can be accessed in
http://www.ucm.es/info/Astrof/UCM\_Survey/UCM/ha\_images.html.}
\end{figure}

We show H$\alpha$ and Cousins-$R_C$ images of some example galaxies in
Figure~\ref{some}. 

\slugcomment{Please, plot this figure in one column format}
\placefigure{some}
\clearpage
\clearpage

\section{Results}
\label{results}

\subsection{Spectra vs. imaging}
\label{specvsima}

One of the main goals of the present work was to determine the amount
of H$\alpha$ emission falling outside the slit employed in the
spectroscopic follow-up of \citet{1996A&AS..120..323G}. The issue
becomes very important when spectroscopic data are used to
characterize extended objects. Different approaches have been followed
when facing the problem, ranging from aperture corrections based in
broad-band fluxes \citep{1995ApJ...455L...1G} to the absence of any
correction \citep{astro-ph/0204055K}.

First, we compared spectroscopic and imaging observations by
simulating the long-slit in the H$\alpha$ frames. Artificial slits
were simulated using the information about position angles and widths
given by \citet{1995PhDT...JGM}. The centers were chosen as the
intensity peaks in the broad-band images, similarly to what was done
in the spectroscopic campaigns. The measurements were repeated 8--10
times with small random variations of the center position and angle,
and then averaged to obtain the most probable value.  The average
values of the spectroscopic fluxes and those measured from the
simulated slits are identical (the average difference is lower than
1\%), although the median fluxes are 5\% larger in the case of the
simulated slits. This is probably related to the fact that the
simulated apertures were more extended along the slit than those used
by \citet{1995ApJ...455L...1G}. Still, individual differences are
reasonably below typical spectroscopic uncertainties.

The integrated H$\alpha$ flux for each object was measured using
polygonal apertures enclosing the entire galaxy and also a curve of
growth built with circular apertures. The estimated errors include
Poisson statistics, uncertainties in the sky determination, the
standard deviation of the photometric zero-points, and the errors
associated with the continuum subtraction.

\begin{figure}
\plotone{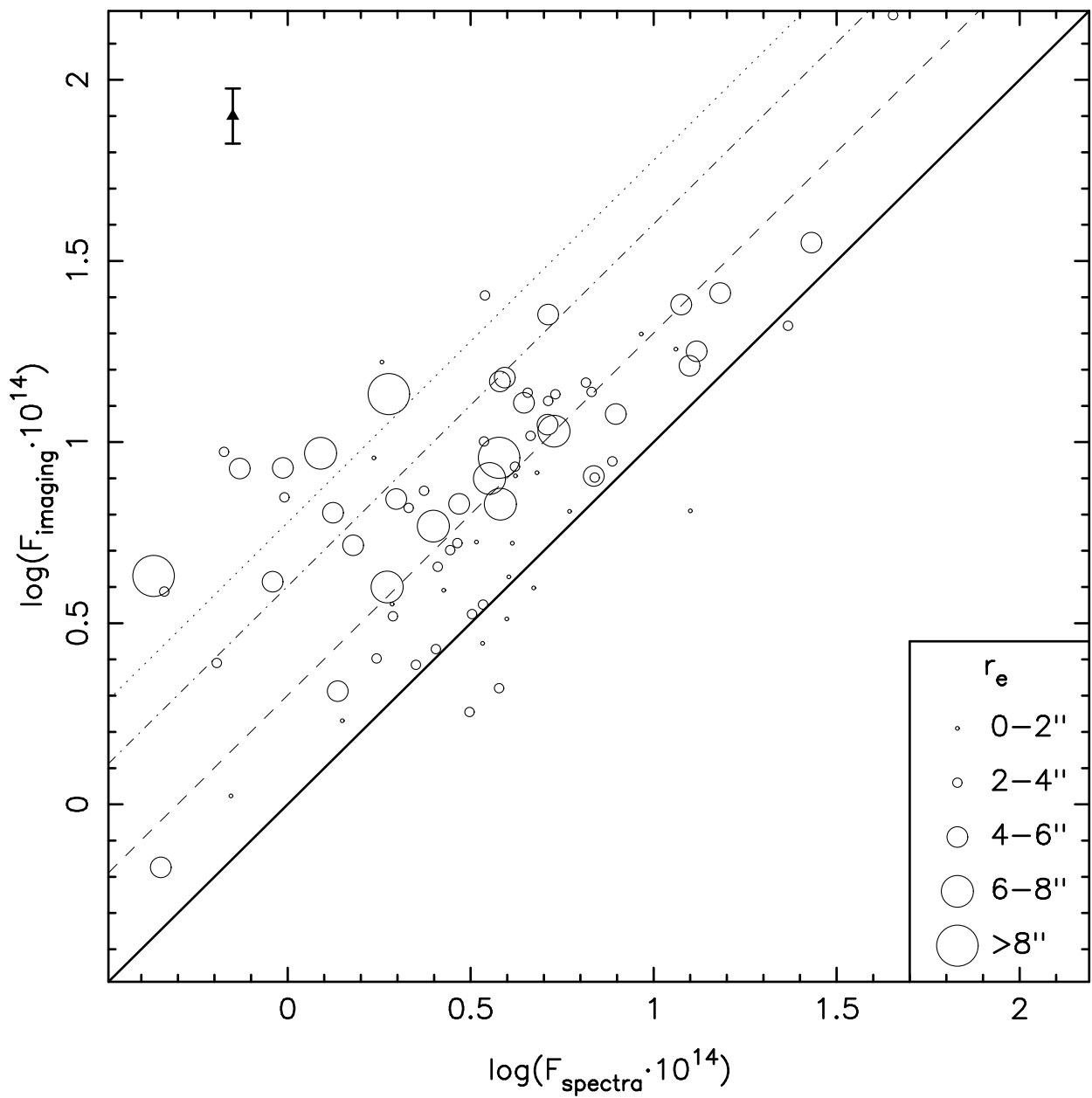}
\figcaption{\label{coraper} Comparison of the H$\alpha$ fluxes determined 
for the spectroscopy and 
those obtained from the integrated emission measured in the images(in
erg~s$^{-1}$~cm$^{-2}$). Symbol sizes are related to the effective
radii in the $B$-band (see legend). The lines correspond to
$F_\mathrm{imaging}/F_\mathrm{spectra}$ ratios of 1, 2, 4 and 6. The
average error in the H$\alpha$ fluxes derived from the images is shown.}
\end{figure}

Figure~\ref{coraper} shows the comparison between the H$\alpha$ fluxes
measured in the spectra and the total H$\alpha$ emission in the images
(decontaminated from [\ion{N}{2}] emission). Symbol sizes are related
to the effective radii in the $B$-band (in arcsec). On average, the
spectroscopic follow-up of the UCM Survey galaxies detected one third
(1/3.0) of the total emission-line flux due to the finite width or a
improper positioning of the slit.  Due to the asymmetry in the
distribution, the corresponding median factor is one half (1/2.0). As
expected, the agreement between photometric and spectroscopic fluxes
is better among the compact objects.

\placefigure{coraper}
\clearpage
\clearpage

Our data also indicates that the global $EW(\mathrm H\alpha)$ values
determined from the imaging data are, on average, a factor $1.8\pm1.2$
smaller than the ones obtained with the long-slit spectroscopy.  This
value is independent of the galaxies' spectroscopic type. Similar
differences have been observed in normal spirals
\citep{1983AJ.....88.1094K}. This can be understood because the
$EW(\mathrm H\alpha)$ is related to the ratio of newly-born stars
(responsible for the emission lines) to the evolved underlying
population, and thus we expect this quantity to be large for galaxies
with relatively important bursts of recent star formation, specially
at very young ages.  Since the spectroscopic observations tend to
concentrate on the brightest star-forming knots, it is not surprising
that the $EW(\mathrm H\alpha)$'s derived from the spectra are higher
than the galaxies' global values.  In addition, the UCM sample is
dominated by objects with bright nuclear starbursts
\citep{1996A&AS..120..323G}, and again we expect the $EW(\mathrm H\alpha)$
measured for the entire galaxy to be smaller than the spectroscopic
values. The exception to this are the galaxies for which the long-slit
was placed in the galaxy center and missed bright HII regions in the
disk.

\subsection{Concentration and size properties}
\label{concen}

The UCM Survey galaxies present enhanced star formation when compared
with `normal' quiescent spiral (or lenticular) galaxies
\citep{2000MNRAS.316..357G,2003MNRAS.338..525P}. Several 
triggering mechanisms are usually postulated for the activation of massive
star formation: tidal forces, mergers, supernova winds, etc. These
processes are more efficient in the central nuclear zones of the
galaxies, where the potential well is deeper and higher gas densities
are found (see \citealt{1998ARA&A..36..189K},
\citealt{2001cksa.conf..475C}, and references therein). Gas stability
criteria would also favour violent star formation events at low radial
distances \citep{1972ApJ...176L...9Q,1989ApJ...344..685K}. Therefore,
it is not surprising that 57\% of the entire UCM sample was
spectroscopically classified as {\it starburst nuclei} galaxies. These
objects are supposed to be experiencing a massive burst characterized
by a high metallicity and low excitation
\citep{1996A&AS..120..323G} and placed in the inner parts of the
galaxy. Another 32\% of the entire survey are HII-{\it like} galaxies
presenting low metallicities and high excitations. Their star
formation could be more extended and/or not linked with the
galactic nucleus.

\begin{figure} \plotone{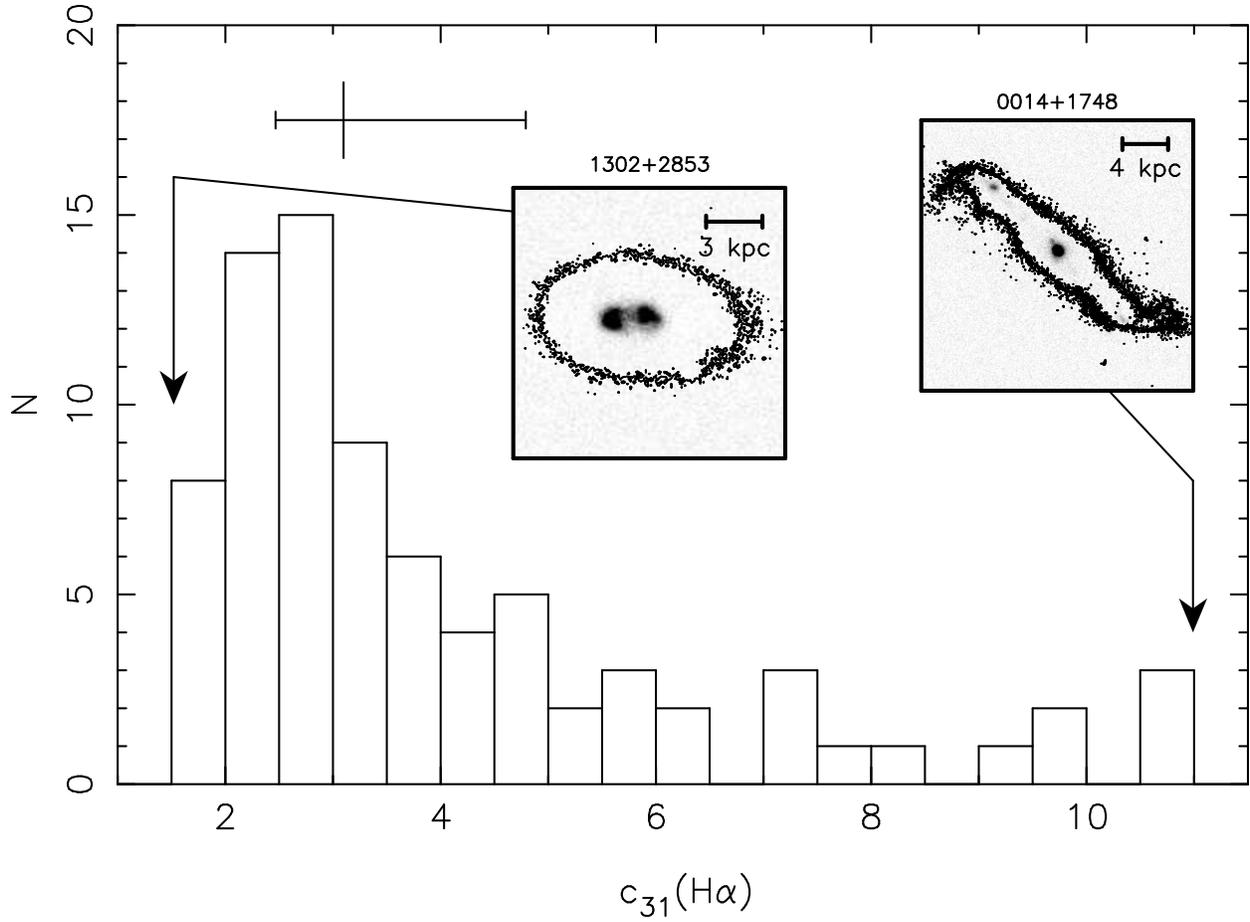}
\figcaption{\label{c31}Distribution of the $c_{31}$ concentration index  for
the UCM sample. The median value and quartiles are shown by the error
bar. The H$\alpha$ images for two extreme cases are plotted in
gray-scale. Overimposed on them are the contour for the pixels with a
surface brightness of $23.5$\,mag~arcsec$^{-2}$ in the $R_C$
image. North is up and east is left. }
\end{figure}

In this section and the next one, we will study the location of the starburst
knots. First, we will focus on the concentration properties of the current star
formation in the UCM Survey galaxies. We have calculated the concentration
index $c_{31}$ for the H$\alpha$ emission. This index was defined by
\citet{1990A&AS...83..399G} as the ratio between the radii of the aperture
containing 75\% and 25\% of the total flux. We have used isophotal
apertures centered in the maximum of the broad-band
images. Figure~\ref{c31} shows the distribution of $c_{31}$ for our
sample of galaxies. The median value and quartiles for the entire
sample are indicated. A wide range of concentrations are observed. The
peak of the distribution is found at $c_{31}=2$--$3$, corresponding to
galaxies with extended star formation. These values are typical of
disk-dominated objects when the morphological classification is based
on broad-band concentration indices
\citep{1990A&AS...83..399G,1996A&AS..120..385V,2001A&A...365..370P}.  Our
sample also includes rather concentrated galaxies with values of $c_{31}$ up to
10.

We must be cautious when interpreting the concentration indices
obtained in H$\alpha$ images and their possible links with nuclear and
disk star formation.  High concentrations may be found in galaxies
with no nucleus or bulge at all but presenting one very bright and
intense burst (e.g, BCDs). In contrast, a single nuclear burst
occupying all the bulge with a flat surface brightness profile may
present a rather low $c_{31}$ value. A more appropriate parameter for
the discussion on nuclear and disk emissions will be used in
Section~\ref{nucdis}.

\placefigure{c31} \clearpage  \clearpage

\begin{figure}
\plotone{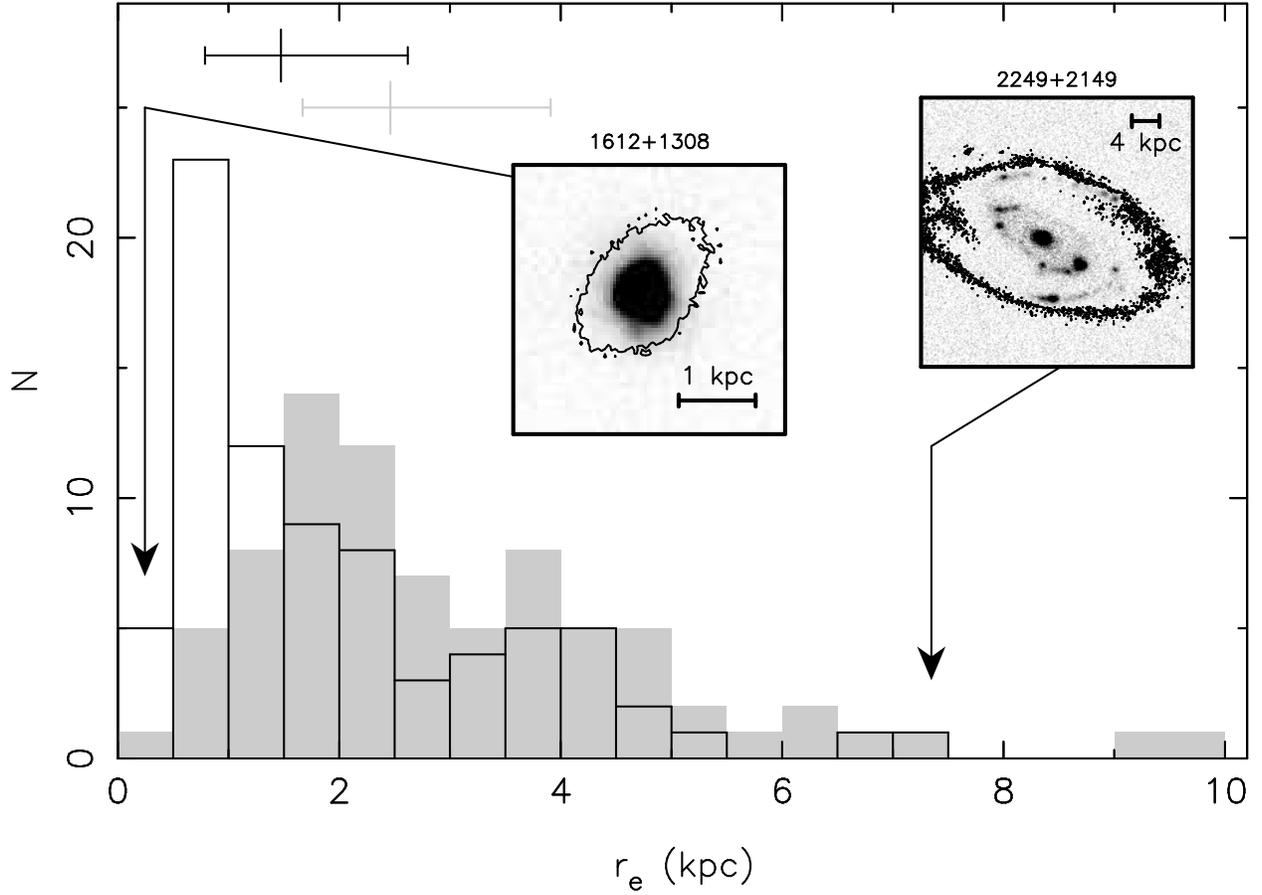}
\figcaption{\label{reff}Distribution of the effective radii (in kpc) of  the
H$\alpha$ (solid-line) and continuum $R_C$ (gray histogram) emission
of the UCM Survey galaxies. As before, the H$\alpha$ images for two
extreme cases are shown (cf.\ figure~\ref{c31}).  }
\end{figure}

Another important point to consider when studying  the concentration of the
emission is the size of the starburst zone. In Figure~\ref{reff} we have
plotted the histogram of the effective radius (i.e., radius containing half of
the light) of the H$\alpha$ emission. The distribution peaks at $\sim 1$\,kpc,
with three quarter of the galaxies presenting effective radii smaller than
$3$\,kpc. The median value is almost identical (ours is 0.02dex larger) to that
found by  \citet{1996ApJ...472..546L} for a sample of edge-on, infrared-warm,
starburst galaxies, and $\sim15$\% larger than the values given by
\citet{2001JApA...22..155C} for a sample of Markarian galaxies. The 
distribution of effective radii in the $R_C$-band is also presented in
this figure. The average is $\sim2.5$\,kpc, which is also in complete
agreement with \citet{1996ApJ...472..546L}. Although we also find a
size-luminosity relationship as the one claimed by these authors, the
scatter is very large. The Spearman's rank-order correlation
coefficient is 0.51 and the two-sided significance level of its
deviation from zero is $2\times10^{-6}$. These size distributions
reveal that, in a significant fraction of the UCM galaxies, the star
formation tends to be more concentrated than the older underlying
stellar population.

We also find a tendency for galaxies with late-type morphologies to
have larger H$\alpha$ effective radii. The median values are: 0.51 kpc
for S0, 0.82 kpc for Sa, 1.38 kpc for Sb, 1.91 kpc for Sc+ and 2.12
kpc for irregulars. The median for the sample of BCDs is 0.53 kpc. The
segregation is also noticeable in the case of spectroscopic types:
disk-{\it like} galaxies present larger values (1.51 kpc) than
HII-{\it like} objects (0.62 kpc).

\placefigure{reff}
\clearpage
\clearpage

\begin{figure}
\plotone{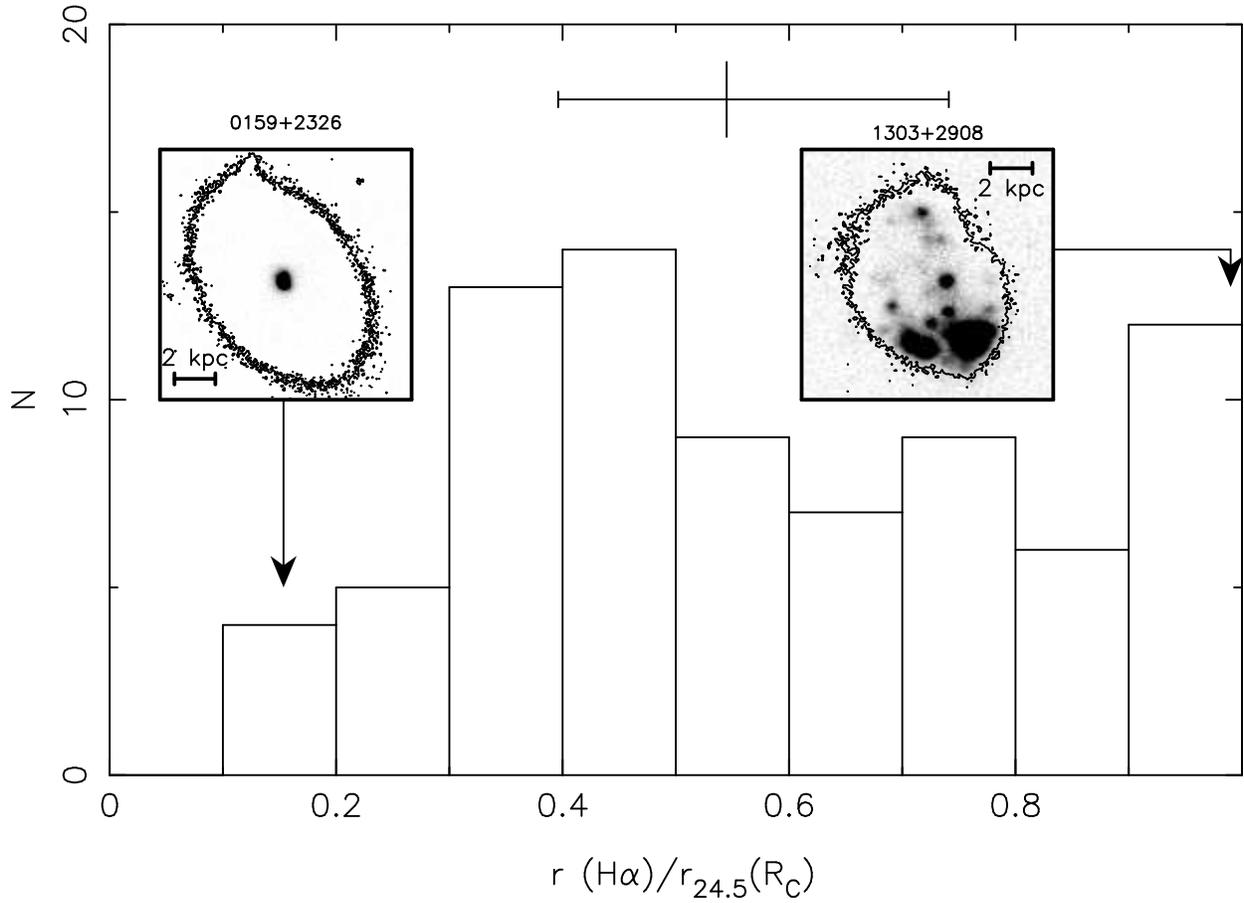}

\figcaption{\label{rhar245}Histogram of the ratio between the sizes of  the
H$\alpha$ emitting region (the radius of the circular region
containing 80\% of the total H$\alpha$ emission) and the entire galaxy
(radius of the $24.5$\,mag~arcsec$^{-2}$ isophote in the $R_C$
image).}

\end{figure}

Figure~\ref{rhar245} presents the histogram of the ratios of the sizes
of the emitting regions and the host galaxies. As a measurement of the
size of the star formation zone we use the radius of the circular
region containing 80\% of the total H$\alpha$ emission. The total size
of the galaxy has been taken as the radius of the
$24.5$\,mag~arcsec$^{-2}$ isophote in the $R_C$ image.  Both regions
have been centered on the maximum-intensity pixel of the broad-band
image. As a consequence, the ratio between the sizes of the H$\alpha$
and continuum emissions is strongly affected by the anisotropy of the
star formation along the galaxy. In particular, this parameter is
strongly affected by the existence of bursts in the outer parts of the
object. The median value of the ratio is $0.5$, i.e., the radius of
the region where there is active star formation is half of the total
optical radius of the galaxy. The values spread from objects where
even the most external zones of the galaxy are involved in the burst
process (e.g., \objectname{UCM1303$+$2908} and
\objectname{UCM1444+2923}) to very concentrated galaxies (e.g.,
\objectname{UCM0159$+$2328} and \objectname{UCM1256$+$2910}). The former are
associated with late-type spirals and irregulars (median values range from 0.5
for S0/Sa to 0.8 for Sc+/Irr,  BCDs and interacting systems). No major
differences are found between the average values of disk and HII-{\it like}
galaxies.

\placefigure{rhar245}
\clearpage
\clearpage

The relative sizes of the star-forming region and the host galaxy may
also be studied by comparing the area of the H$\alpha$ emitting zone
and the total extension of the object. This parameter is not affected
by the anisotropy of the star formation. The area of the burst has
been calculated using the procedure explained in
Section~\ref{hiidiff}, and the extension of the galaxy has been
assumed to be the area inside the $24.5$\,mag~arcsec$^{-2}$ isophote.
The distribution for the entire sample is flat, with a median value of
0.5 and quartiles at 0.3 and 0.9. Again, a correlation with Hubble
type is observed (from 0.3 for S0/Sa to 0.7 for Sc+/Irr and 1.0 for
BCDs). On average, disk-{\it like} galaxies harbor a star-forming
event covering $\sim40$\% of the area of the galaxy, while $\sim70$\%
of the area is covered by star formation in the case of HII-{\it like}
objects.

\begin{figure}
\plotone{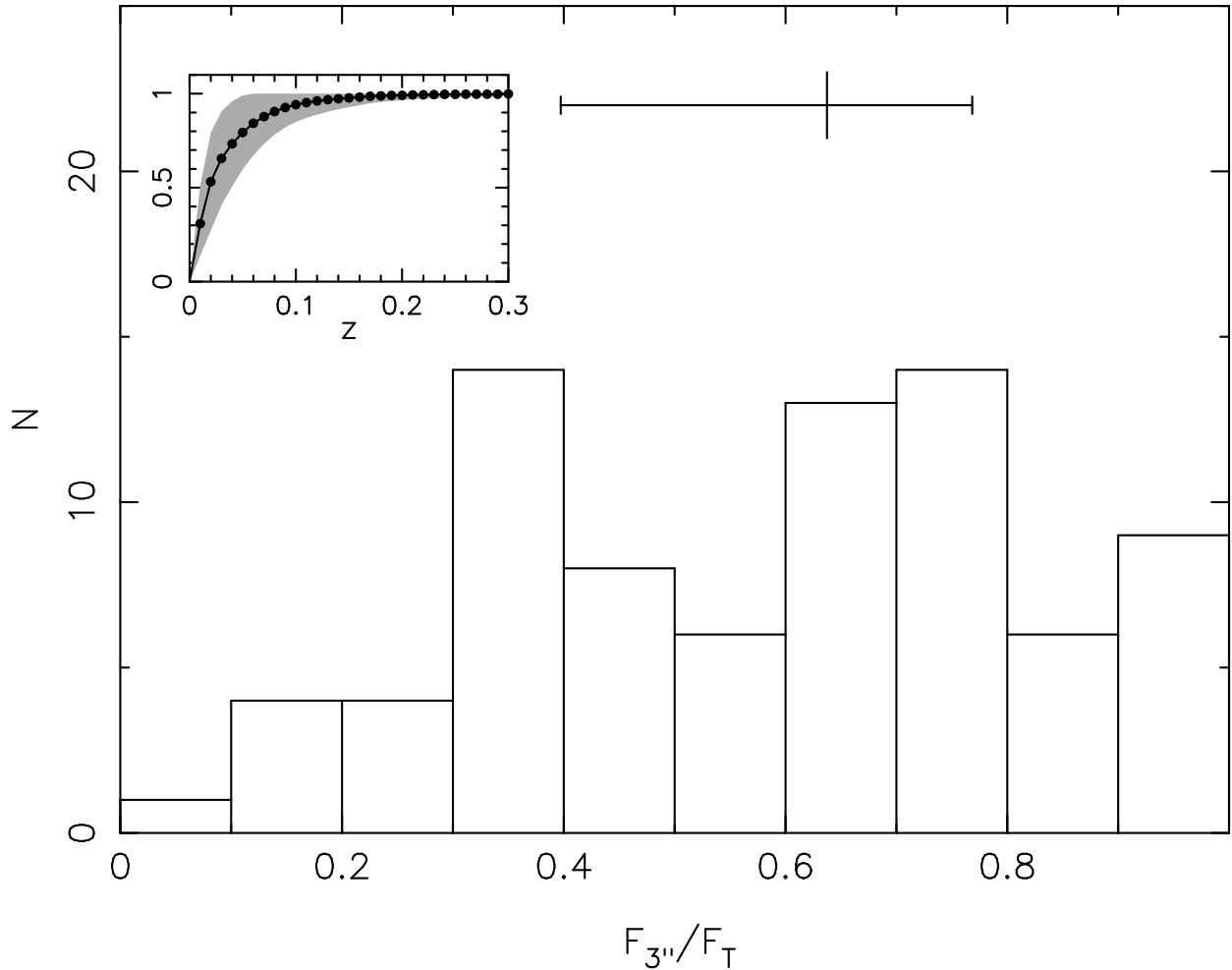}
\figcaption{\label{three}Comparison of the H$\alpha$ emission 
within a circular aperture of $3\arcsec$ (typical of wide field and
all sky surveys such as 2MASS or SDSS) to the total flux for the UCM
galaxies. The inset in the top-left corner shows the change with
redshift of the flux ratio in a $3\arcsec$ circular aperture. The
values for the entire sample have been calculated by
artificially-redshifting the UCM galaxies. Mean values, standard
deviations (shaded area) and a polynomial fit are shown.}
\end{figure}

Our study of the sizes of the star-forming regions in local galaxies  is of
direct relevance to the interpretation of spectroscopic data from large galaxy
surveys such as the Two Micron All Sky Survey
\citep[2MASS,][]{2000AJ....119.2498J} or the Sloan Digital Sky Survey
\citep[SDSS,][]{2002AJ....123..485S}.  In these surveys, the spectroscopic data
is obtained using multi-object spectrographs which utilize multi-slit
masks or fibers. These slits and fibers have a finite size and force
to consider aperture corrections. Our imaging study may be used to
estimate the importance of these corrections. For example,
Figure~\ref{three} shows the distribution of the ratios of the
H$\alpha$ flux within an aperture of $3\arcsec$ (used by the SDSS) to
the total H$\alpha$ flux. A large spread exists in the fractional
coverage of the UCM galaxies by this aperture size. The median of the
distribution reveals that the SDSS fibers would miss typically 40\% of
the total H$\alpha$ emission of UCM-like galaxies, and that more than
half of the flux would be missed for 40\% of our sample.

The inset in the top-left corner of Figure~\ref{three} shows the
change with redshift of the observed-to-total H$\alpha$ flux ratio in
a $3\arcsec$ circular aperture.  We have artificially redshifted the
UCM galaxies from their measured redshifts up to $z=1$ in $\Delta
z=0.01$ steps, and estimated the fraction of the fluxes that would be
detected inside a $3\arcsec$ spectroscopic fiber (or slit) for each
redshifted object. The calculated points in the top-left panel of
Figure~\ref{three} are the averages for the entire sample.  They have
been fitted with a polynomial curve. This plot reveals that only
beyond redshift $\sim0.1$ the $3\arcsec$ aperture contains more than
95\% of the total flux. For redshifts below this limit, corrections to
the H$\alpha$ aperture fluxes are non-negligible.

\placefigure{three}
\clearpage
\clearpage

\subsection{Nuclear vs. disk emission}
\label{nucdis}

The relative contribution of the nuclear and the disk components to
the total H$\alpha$ luminosity will be studied in this section. As we
mentioned in Section~\ref{concen}, concentration indices may lead to
errors when analyzing the location of the starbursts if a bulge-disk
characterization is not performed previously. Bulge-disk decomposition
for the UCM Survey galaxies was carried out in the Gunn-$r$ and
Johnson-$B$ bands by \citet{1996A&AS..120..385V} and
\citet{2001A&A...365..370P}, respectively. Using the results for the
red filter, we have considered:

\begin{equation}
\label{defin}
F_\mathrm T^\alpha=F_{\mathrm {bulge}}^\alpha+F_{\mathrm
{disk}}^\alpha=\frac{\int_{0}^{\infty}F^\alpha(r)\cdot I^r(r)\cdot
r\cdot dr}{\int_{0}^{\infty}I^r(r)\cdot r\cdot dr}
\end{equation}

\noindent where $F_T^\alpha$ is the total H$\alpha$ flux, $F^\alpha(r)$ is 
the radial dependent H$\alpha$ flux, $I^r(r)=I^r_{\mathrm
{bulge}}+I^r_{\mathrm {disk}}$ is the intensity in the Gunn-$r$
filter, and the integral extends over all radii.

Identifying the nuclear component with the bulge, the H$\alpha$ flux
$F_{\mathrm {bulge}}^\alpha$ arising from this area can be calculated
using:

\begin{equation}
\label{defin2}
	F_{\mathrm bulge}^\alpha=\frac{2\pi}{L^r_\mathrm
	T}\cdot\int_{0}^{\infty}F^\alpha(r)\cdot I^r_{\mathrm
	bulge}(r)\cdot r\cdot dr
\end{equation}

\noindent where $L^r_\mathrm T$ is the integrated luminosity in the $r$ 
filter. An analogous expression can be built for the H$\alpha$ flux of
the disk. In \citet{1996A&AS..118....7V}, the bulge intensity profile
was fitted using a \citet{1948AnAp...11..247D} exponential function,
and the disk intensity was fitted by a \citet{1970ApJ...160..811F}
law. In order to calculate the numerator of Equation~\ref{defin2}, the
function $F^\alpha(r)$ was approximated by a series of Chebyshev
polynomials. The integral was calculated using a Gauss-Laguerre
quadrature.

The method described above was used to obtain the ratio between the
bulge and the total H$\alpha$ luminosities --(B/T)(H$\alpha$)--. The
goodness of the bulge-disk decompositions was checked interactively
for each individual object.

The median and standard deviation of the (B/T)(H$\alpha$) ratios of
the sample are $0.26\pm0.26$. This means that a typical UCM galaxy
harbours a star formation event with $\sim30$\% of nuclear
component. Therefore, whereas the spectroscopic classification was
mainly dominated by starburst nuclei
\citep[SBN,][]{1983ApJ...268..602B} galaxies, the images reveal that
simultaneously to the nuclear burst there exists an important amount of recent
star formation occurring in the disk. A segregation of (B/T)(H$\alpha$) values
according to the spectroscopic classification has been observed: disk-{\it
like} objects present a median of $0.33\pm0.26$ and HII-{\it like}
$0.15\pm0.26$. The (B/T)(H$\alpha$) is correlated with morphological type in
the UCM Survey, as some authors have claimed for other samples \citep[see,
e.g.,][]{1983ApJ...272...54K,1998ARA&A..36..189K,1996ApJ...472..546L}.  The
median values are: 0.43 for lenticulars, 0.45 for Sa, 0.36 for Sb, and 0.05 for
late Hubble types (Sc+/Irr) and BCDs.

\subsection{Equivalent width and luminosity}
\label{ewl}

In Section~\ref{specvsima}, we argued that $EW(\mathrm H\alpha)$
provides a measurement of the relative importance of newly-formed
stars with respect to the older population.  Unlike `normal' spiral
galaxy samples, all the spiral galaxies in the UCM sample have median
integrated $EW(\mathrm H\alpha)$ (based on imaging data) of about
$40-50$~\AA. Even more intense star formation is found in irregulars
(the average is $70$~\AA), BCDs ($110$~\AA), and interacting systems
($60$~\AA). The correlation observed between morphological type and
$EW(\mathrm H\alpha)$ for normal galaxies
\citep{1983AJ.....88.1094K,1994ApJ...429..153S,1998AJ....115.1745G} is not 
present in our sample, probably as a consequence of the enhanced star-forming
activity of the UCM galaxies \citep[see Section 3, and figure 1
in][]{2003MNRAS.338..508P}. This is directly related to selection effects of
the survey, biased towards the detection of objects with high $EW(\mathrm
H\alpha)$ \citep[larger than $\sim20$~\AA,][]{1995PhDT...JGM}.

Spectroscopically, disk-{\it like} galaxies present considerable lower
values of the $EW(\mathrm H\alpha)$ ($24$~\AA) than HII-{\it like}
objects ($75$~\AA).  This is directly related to the findings of
\citet{2000MNRAS.316..357G} and \citet{2003MNRAS.338..525P}:   
HII-{\it like} objects have higher SFR per unit stellar mass than
disk-{\it like} galaxies.

\begin{figure}
\epsscale{1.0}
\plotone{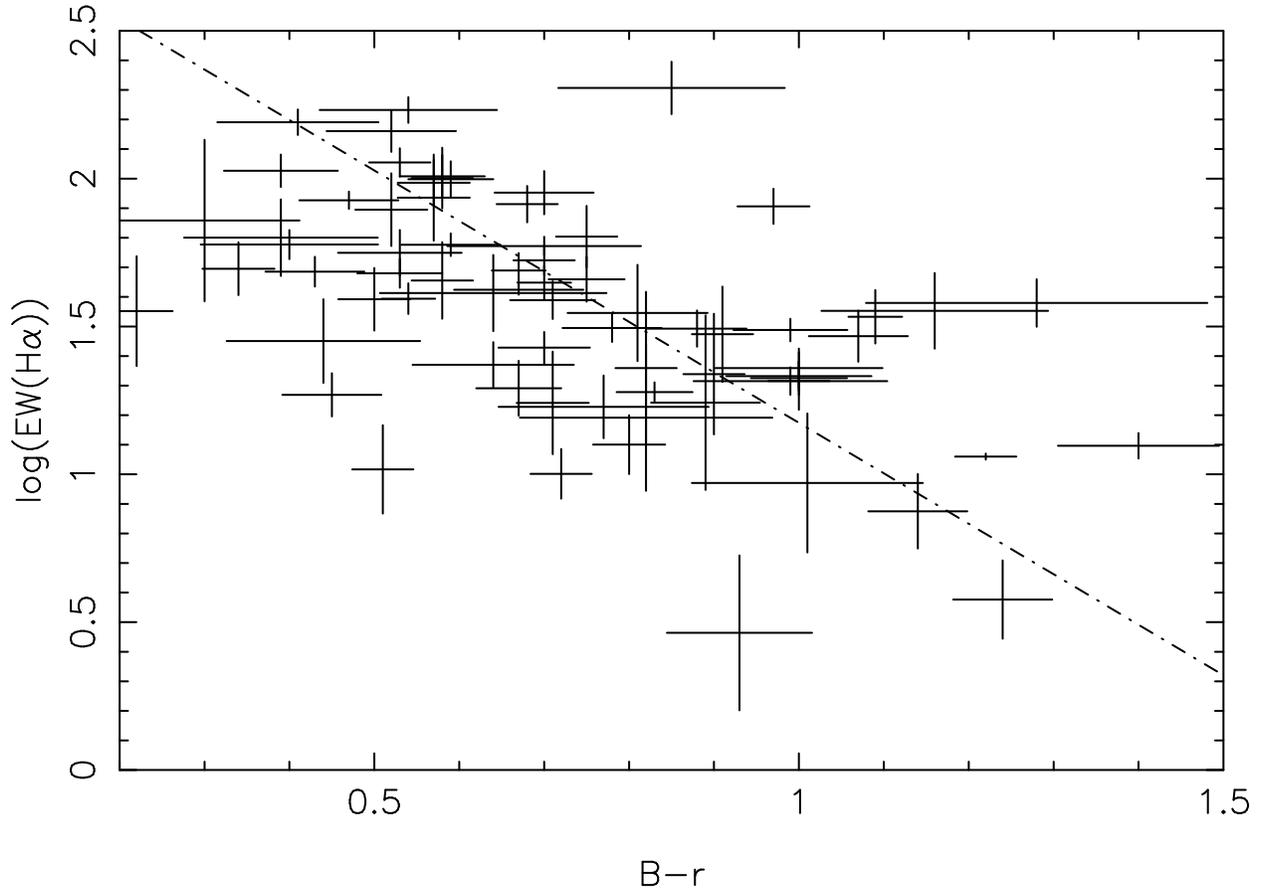}
\figcaption{\label{br_ew} Integrated optical colour ($B-r$) 
vs. imaging equivalent width (\AA) relationship. The best fit is
plotted: $\log(EW(H\alpha))= (-1.707\pm0.091)\times(B-r)+
(2.881\pm0.066)$. The Spearman's rank-order correlation coefficient is
-0.55 and the two-sided significance level of its deviation from 0 is
$4\times10^{-7}$.}
\end{figure}

The integrated H$\alpha$ emission has been found to be strongly
correlated with optical colours
\citep{1983AJ.....88.1094K,1984AJ.....89.1279K,1992AJ....103.1512D}.  This
means that the youngest stars, though not contributing much to the total
stellar mass of the galaxy ($\sim5$\% for the UCM galaxies, cf.
\citealt{2003MNRAS.338..525P}), have a major influence on the integrated
optical colours. The correlation is also observed for the UCM sample and is
plotted in Figure~\ref{br_ew}. The large scatter of this plot is related to
differences in star formation history and  dust extinction.

\placefigure{br_ew}
\clearpage
\clearpage

\begin{figure}
\plotone{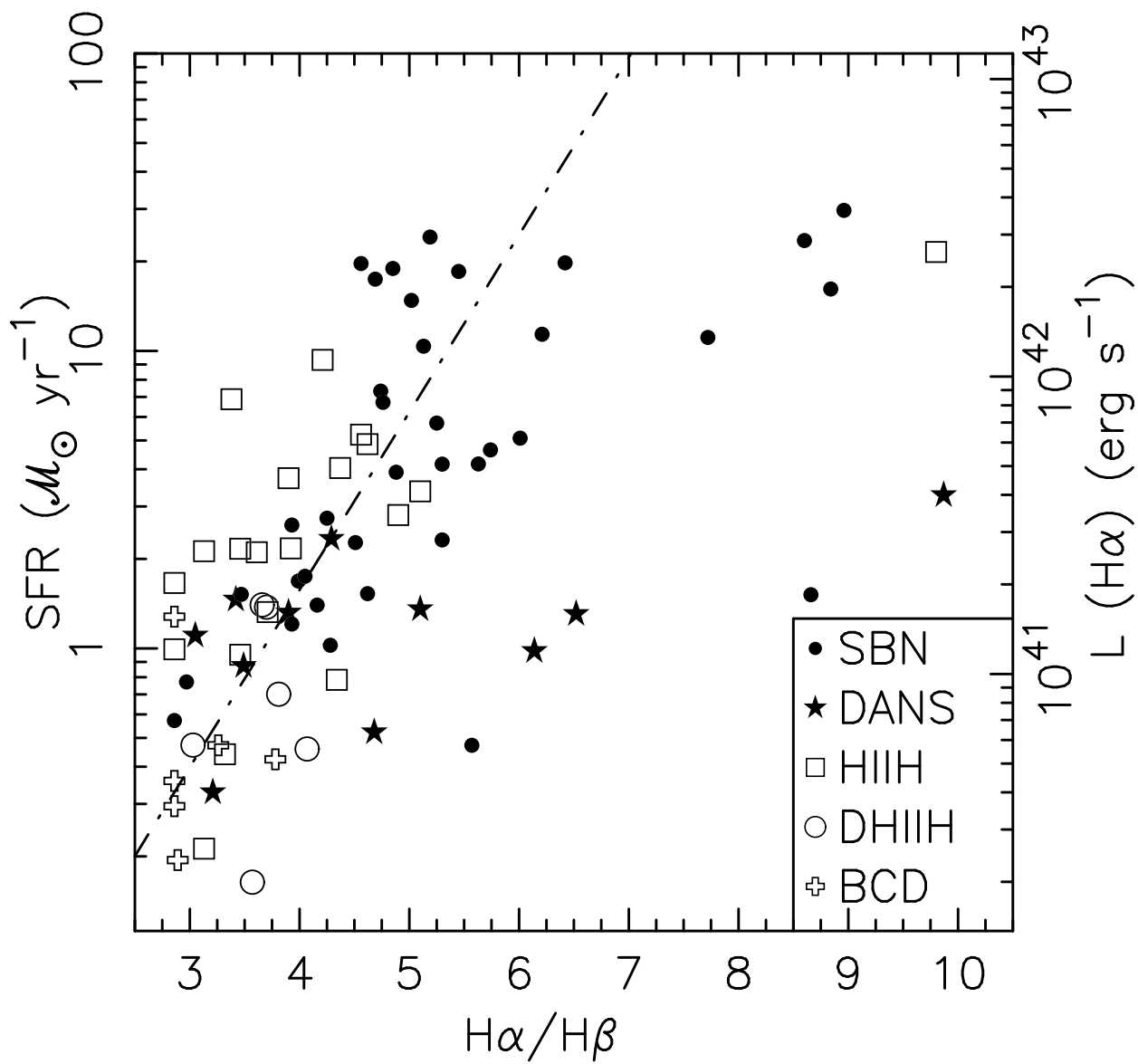}
\figcaption{\label{sfr_hahb} Relationship between extinction and   SFR (or
H$\alpha$ luminosity, as shown on the right-hand side axis).  The best
linear fit to the data, which takes into account observational errors,
is shown (see text).  Different symbols are used for different
spectroscopic types (see legend). Filled symbols correspond to
disk-{\it like} galaxies and open ones to HII-{\it like} objects
\citep[see][ and Section~\ref{sample}]{1996A&AS..120..323G}.}
\end{figure}

As the next step in our study, we have calculated the H$\alpha$
luminosities for the entire sample so that we can compare the
imaging-based results with those obtained from spectroscopy (cf.\
Section~\ref{ro}). But before we carry out such comparison, it is
interesting to see whether the measured luminosities (or the derived
SFRs) correlate with any other observables.  In Figure~\ref{sfr_hahb}
we have plotted the SFR against the $\mathrm H\alpha/\mathrm H\beta$
ratio, a useful extinction estimator.  SFRs have been calculated using
the conversion factor given in \citet{2001ApJ...558...72S}, almost
identical to the one obtained by \citet{1998ARA&A..36..189K}:

\begin{equation}
\label{convfac}
SFR=\frac{L(H\alpha) (erg~s^{-1})}{1.22\times10^{41}} \mathcal M_\sun~\mathrm{yr}^{-1}
\end{equation}

Figure~\ref{sfr_hahb} shows a clear correlation between the derived
SFRs and the extinction.  Objects with larger obscurations have also
larger SFRs. This effect was also noticed in the stellar population
synthesis analysis presented in \citet{2003MNRAS.338..525P}, where not
only high-extincted galaxies showed higher burst strengths (ratio
between the mass of newly-formed stars and the total stellar mass),
but they also seemed to follow different extinction laws than
low-extinction objects.


\placefigure{sfr_hahb}
\clearpage
\clearpage

The correlation between SFR and extinction has also been reported in
previous works
\citep[e.g.,][]{2001AJ....122..288H,2001ApJ...558...72S}. Our fit to
the relationship between the SFR and the Balmer decrement (corrected
by stellar absorption) is, taking into account observational errors:

\begin{equation}
\log(SFR)=(0.597\pm0.021)\times \frac{\mathrm H\alpha}{\mathrm H\beta}+(-2.191\pm0.086)
\end{equation}

The Spearman's rank-order correlation coefficient is 0.67 and the
two-sided significance level of its deviation from 0 is
$2\times10^{-11}$. The derived slope is much larger than the one given
in \citet{2001ApJ...558...72S} and \citet{2001AJ....122..288H}. It is
interesting to notice that the correlation seems to saturate for very
reddened objects ($\mathrm H\alpha/\mathrm H\beta\gtrsim7$,
corresponding to $E(B-V)\gtrsim0.9$), and therefore the derived
regression is not valid for very high extinction values. This behaviour
was not observed by \citet{2001ApJ...558...72S} since their
ultraviolet-selected nearby galaxy sample lacks very highly-obscured
objects. This is not surprising: the selection at ultraviolet (UV)
wavelengths is strongly biased against highly obscured objects.  Note
also that if we only consider the objects in our sample with Balmer
decrements lower than 5.5 (i.e., low-to-moderate extinction), the
correlation we find becomes very similar to the one found by
\citet{2001ApJ...558...72S}.

Most of the objects which do  not follow our observed correlation are disk-{\it
like} galaxies. Note also that the DANS (Dwarf amorphous nucleated spheroidals,
\citealt{1989ApJS...70..479S}) seem to follow  a different relation from the
rest of the objects, presenting relatively low and almost constant
SFRs ($\sim$1~$\mathcal{M}_\sun$~yr$^{-1}$) for all obscurations. This
spectroscopic class, the low luminosity version of SBN objects, has
already shown significant differences in other properties such as
their stellar mass and SFR per stellar mass
\citep{2003MNRAS.338..525P}.

The breakdown of the SFR--extinction relationship at high $\mathrm
H\alpha/\mathrm H\beta$ could be a symptom of the fact that,  at large
obscurations, extinction indicators such as  the Balmer decrement or the slope
of the UV continuum fail to correct the observed luminosities
\citep{2002ApJ...577..150B}.  This is also supported by observations of some
ultraluminous infrared galaxies, where the UV slope indicates a lower
extinction than that derived from the UV to far infrared (FIR) ratio
\citep{2002ApJ...568..651G}.  Observations of the UCM sample in the FIR would
be necessary to ascertain whether, by using  the observed $\mathrm
H\alpha/\mathrm H\beta$, we significantly underestimate the extinction  (and
thus the SFR) in some heavily-obscured star-forming regions within the
galaxies. It could well be  that some star-formation occurs in regions where no
H$\alpha$ photons are able to escape, and therefore such star-formation would
be missing in an  optically-based SFR census.

\subsection{Star-forming knots and diffuse emission}
\label{hiidiff}

This section presents the results concerning the analysis of the
spatial distribution and intensity of the star formation in the UCM
galaxies. This study has two main goals: (1) to segregate the
star-forming knots from the regions with a more diffuse ionized
gaseous component (DIG), and (2) to characterize the regions with
active star formation.

\subsubsection{Detection method}
\label{method}

In order to separate the distinct star-forming knots from the diffuse emission,
we have followed a similar method to the one used by
\citet{1996A&A...307..735R} and \citet{2001PhDT...Zurita} to detect HII regions
and study their luminosity functions in nearby galaxies.  The main
difference is that, due to their higher distances, the angular sizes
of the UCM galaxies are much smaller than those of the galaxies
studied by these authors. Thus, the average spatial resolution element
of our H$\alpha$ imaging study (i.e., the diameter of the typical
seeing disk, $\sim1$\arcsec) corresponds to $\sim0.5$~kpc. This value
is similar to the diameters of the largest HII regions found in nearby
galaxies
\citep[$\sim400$~pc,][]{1999ApJ...519...55Y,2001PhDT...Zurita}. Therefore,
the star-forming regions detected in our sample correspond to large
star-forming complexes rather than normal HII regions. Moreover, the
UCM objects present enhanced star formation in comparison with normal
quiescent galaxies. This behavior also has an effect on the size and
intensity of the starbursts.

The detection method of the star-forming knots deals with four different
issues:

\begin{itemize}

	\item[1.--]The global limit for the detection of any feature
	was set to 2 times the noise level of the sky background.

	\item[2.--]The criterion for distinguishing an individual
	star-forming knot from its neighbors was that every local
	maximum with no other maxima closer than 2 pixels was
	considered as an individual knot. This separation was chosen
	so that the diameter of any knot would be at least as large as
	the average diameter of the seeing disk in our frames
	($\sim1\arcsec=5$~pixels).

	\item[3.--]The technique used to decide which pixels belong to
	each region when there are other regions nearby is described
	in \citet{2001PhDT...Zurita}. The technique is based on the
	same principles used by the SExtractor software
	\citep{1996A&AS..117..393B} to de-blend objects.

	\item[4.--]The detection limit for the boundaries (faintest
	parts) of the star-forming regions was chosen to coincide with
	the upper intensity limit for the diffuse emission (see
	ahead). The value we used was
	$1.32\times10^{-16}$~erg~s$^{-1}$~cm$^{-2}$~arcsec$^{-2}$. The
	limit corresponds to an emission measure\footnote{The emission
	measure $EM$, whose units are pc~cm$^{-6}$, is defined as
	$EM=\int_{0}^{d}N_e^2~ds$, where $N_e$ is the electron density
	and the integral extends to the line of sight. For an electron
	temperature of $T=10^4$~K,
	$EM=4.908\times10^{17}$~I(H$\alpha$), where I(H$\alpha$) is
	the intensity in erg~s$^{-1}$~cm$^{-2}$~arcsec$^{-2}$.}
	$EM=65~$pc~cm$^{-6}$ . The choice of this value is based on
	several studies of the diffuse ionized medium in spiral
	galaxies (for example,
	\citealt{1996AJ....112.2567F,1996AJ....111.2265F} use a cutoff
	value of $80$~pc~cm$^{-6}$, and \citealt{1996AJ....112.1429H}
	use $50$~pc~cm$^{-6}$).

\end{itemize}

\subsubsection{Star-forming knots}

In the 79 galaxies included in the present study, we have detected 816
distinct star-forming knots. The number of regions found in each
galaxy ranges widely, from one single region per galaxy to up to 87
regions for one galaxy (\objectname{UCM2325$+$2318}, a luminous
infrared galaxy, \citealt{2000ApJ...533..682C}). The median value is
$\sim8$ regions per galaxy.  We detect more regions in late-type
spirals and interacting systems (on average, 12 regions in Sc+
galaxies, 18 in Irr and 45 in interacting systems) than in early-type
spirals (5--8 for S0/Sb) and BCDs (1 region).

Interestingly, no strong correlation is observed between the number of
regions detected within an object and the distance to the galaxy.  The
Spearman's rank correlation coefficient for these parameters is 0.2,
with a two-sided significance level of its deviation from 0 of
$4\times10^{-2}$.  This indicates that, with our definition of
star-forming knots, spatial resolution is not a major factor in the
number of star-forming knots detected for the range of distances
covered by the UCM galaxies. In contrast, the number of regions and
the apparent size of the entire galaxy (taken as the radius of the
$24.5$\,mag~arcsec$^{-2}$ isophote) do show some correlation: larger
galaxies tend to have more star-forming regions. The Spearman's rank
correlation coefficient of these quantities is 0.48 and the two-sided
significance level of its deviation from 0 is $7\times10^{-6}$. In
addition, those galaxies with large numbers of regions are also more
luminous than objects with fewer knots of star formation.

The median radius of the star-forming knots in the UCM galaxies is
0.5~kpc, which is approximately twice the average resolution
(cf. Section~\ref{method}).  The smallest detected knot has a radius
of $\sim100$~pc. The starburst complexes can be as large as 4~kpc. The
most extreme cases are found in
\objectname{UCM1513$+$2012} and \objectname{UCM2250$+$2427}, two very disturbed
galaxies. 

\begin{figure}
\plotone{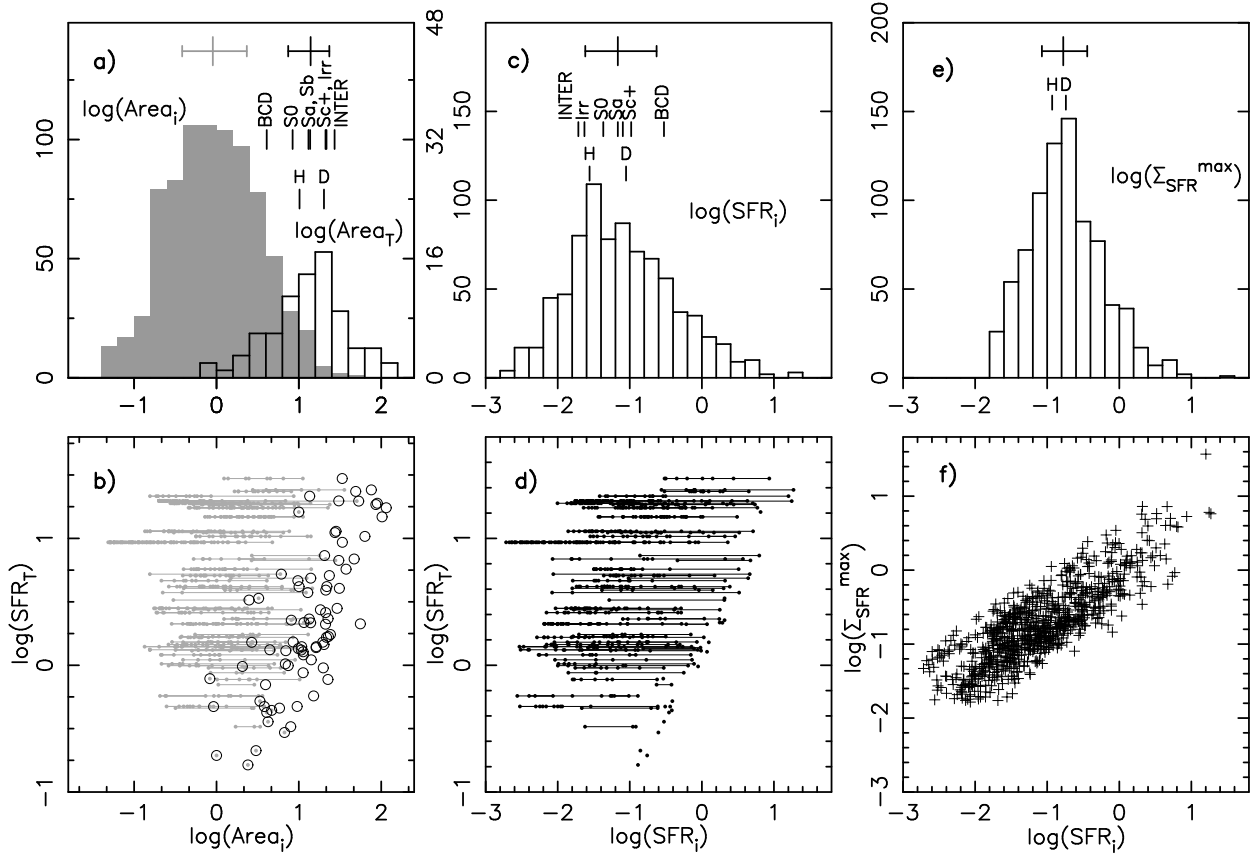}
\figcaption{\label{varhii}Frequency distributions and correlations of 
the properties of the star-forming knots detected in the UCM
sample. Areas are in kpc$^2$, SFRs in $\mathcal{M}_\sun$~yr$^{-1}$,
and $\Sigma_{\mathrm{SFR}}^{\mathrm{max}}$ in
$\mathcal{M}_\sun$~yr$^{-1}$~kpc$^{-2}$. Panels show: a) Histograms of
the area of the individual star-forming knots (Area$_\mathrm i$, in
gray) and the total extension of the H$\alpha$ emission (Area$_\mathrm
T$). Median values and quartiles of the entire distributions are
marked by the error bars. The median values for the different
morphological and spectroscopic classes are also marked. b)
Relationship between the global SFR and the extension of the H$\alpha$
emission. Data for the individual knots are plotted with gray
points. A horizontal gray line joins the knots belonging to the same
galaxy (when more than one knot have been detected). Data referring to
the integrated area of the H$\alpha$ emission are depicted with open
circles. c) Distribution of the SFRs of individual star-forming knots
within the UCM galaxies. Median values and quartiles of the entire
distribution and for each morphological and spectroscopic types are
marked. d) Relationship between the global SFR and the SFR of the
individual knots. A horizontal line joins the knots belonging to the
same galaxy (when more than one knot have been detected). e)
Distribution of the maximum surface density of SFR
($\Sigma_\mathrm{SFR}^{\mathrm{max}}$; see text). Median values and
quartiles of the entire distribution and different spectroscopic types
are marked. And f) $\Sigma_\mathrm{SFR}^{\mathrm{max}}$ vs. SFR$_i$
for the detected star-forming knots in the UCM sample.}
\end{figure}

\slugcomment{Please, plot this figure in one column format}
\placefigure{varhii}
\clearpage
\clearpage

Using the conversion factor given in Equation~\ref{convfac} we have
also calculated global SFRs (SFR$_\mathrm T$), average SFR surface
densities ($\Sigma_{\mathrm{SFR}}$), and maximum SFR surface densities
($\Sigma_{\mathrm{SFR}}^{\mathrm{max}}$, the SFR surface density of
the pixel with the highest intensity) for each of the regions.

The gray histogram in Figure~\ref{varhii}(a) shows the frequency
distribution of the surface areas of the individual star-forming
knots. The projected areas range from 0.05 to 50~kpc$^2$. The
distribution peaks at $\sim1$~kpc$^2$, with a median value of
0.9~kpc$^2$. The frequency distribution of the total area of the
H$\alpha$-emitting regions in each galaxy is also shown (empty
histogram).  This distribution presents a peak at
$\sim14$~kpc$^2$. Median values for the different morphological and
spectroscopic types are marked. A trend is observed in the sense that
later Hubble types have larger active star-forming areas.  As expected
for their dwarf nature, BCDs are characterized by the small extension
of the burst zone. On the contrary, the objects that were classified
as highly-disturbed interacting systems present the most extended
bursts (the median is $27$~kpc$^2$). From a spectroscopic point of
view, disk-{\it like} galaxies show emitting total areas which are
twice the size of those of HII-{\it like} objects.

Figure~\ref{varhii}(b) shows the correlation between the total area
covered by the H$\alpha$ emission and the SFR of the galaxy (open
circles).  Data for the individual regions in each galaxy is also
shown as dots joined by a line.  The total area covered by
star-formation is each galaxy is reasonably well correlated to the
total SFR, and the logarithmic slope of the relation is close to
1. This indicates that the range in global SFR surface densities is
relatively narrow for all the galaxies in the UCM sample.  Indeed, the
median value of the total SFR per unit star-forming area in the UCM
sample is $0.15\,\mathcal{M}_\sun$~yr$^{-1}$~kpc$^{-2}$, with an rms
of only $0.4$dex.  Note that since these numbers have not been
corrected for inclination, the intrinsic scatter is probably smaller.
Interestingly, we also find the most extended individual knots in the
most actively star-forming galaxies (see discussion of
Figure~\ref{varhii}(d) below).

Figure~\ref{varhii}(c) shows the frequency distribution of the SFRs
estimated for the individual regions. Median values for the global
distribution and for each morphological and spectroscopic types are
shown. There is, again, a clear trend with morphology for the spiral
galaxies: later types have more intense star-forming knots than
earlier spirals.  BCDs are segregated from the rest of the galaxies
since they tend to have the brightest star-forming regions. Somewhat
surprisingly, irregular and interacting galaxies seem to have
star-forming regions with lower individual SFRs, but since, as we
found before, they tend to have large numbers of them, their global
SFRs can still be high. The star-forming knots in disk-{\it like}
galaxies are also more intense (by a factor of a few) than those found
in HII-{\it like} objects.  This seems to suggest two different
star-formation regimes operate in `quiescent' objects (e.g., normal
spirals) and disturbed/interacting objects.

Figure~\ref{varhii}(d) shows that high integrated H$\alpha$ emission
in a galaxy is accompanied by the existence of more intense individual
star-forming regions.  This trend may be explained in part by an
statistical effect. If the frequency distribution shown in
figure~\ref{varhii}(c) is sampled randomly in each galaxy, the most
intense star-forming knots would tend to appear in galaxies with large
SFRs. In some cases, galaxies with low SFR will be completely
dominated by one single burst. However, if two different
star-formation regimes operate in disturbed and `quiescent' galaxies,
this simple statistical explanation would be complicated by the
possible existence of different frequency distributions for the
different star-formation modes.  Moreover, simple statistics do not
provide the whole story. In a significant fraction of the galaxies
where multiple star-forming knots are found, the global star formation
is completely dominated by one of them (often the central/nuclear one;
cf.\ section~\ref{concen}). Obviously, this also contributes to the
observed correlation.

Finally, figure~\ref{varhii}(f) presents the tight correlation
existing between the SFR of individual star-forming knots and their
maximum $\Sigma_{\mathrm{SFR}}$. A similar behaviour is observed
between the average $\Sigma_{\mathrm{SFR}}$ of each knot and the total
SFR, though a larger scatter is found. The values of the maximum
$\Sigma_{\mathrm{SFR}}$ that we find are lower than the maximum limit
found for starbursts by \citet{1997AJ....114...54M}, which is
consistent with the fact that most of our galaxies are not extreme
starbursts.  We observe no correlation between
$\Sigma_\mathrm{SFR}^{\mathrm{max}}$ (or $\Sigma_\mathrm{SFR}$) and
morphological type.


Summarizing the results shown in figure~\ref{varhii}, the most intense
star-forming knots present the highest SFR surface densities and
largest sizes.  These properties eventually translate into a larger
global H$\alpha$ emission for the entire galaxy. High-resolution
studies have also found this correlation between the global SFR and
the number and mass of the individual star-forming knots \citep[see][
and references therein]{1998ARA&A..36..189K}.  Even though these
trends apply to the vast majority of the UCM galaxies, we also find
suggestive evidence that a different star-formation mode could be
operating in the irregular/interacting galaxies in our sample.

\subsubsection{Diffuse H$\alpha$ emission}

We have followed two different approaches when separating the diffuse
ionized gas (DIG) emission from the star-forming knots. First, we used
an absolute surface brightness cutoff value that corresponds to the
faintest emission that we considered to be part of a star-formation
knot in the previous section
($1.32\times10^{-16}$~erg~s$^{-1}$~cm$^{-2}$~arcsec$^{-2}$). Such an
approach does not deal with spatial variations of the DIG intensity
within a galaxy or from one object to another.  For that reason, we
have also used a more elaborated detection technique that takes into
consideration distinctive properties of the DIG such as its
diffuseness or its morphology. Following \citet{2001PhDT...Zurita}, we
have built gradient maps of the H$\alpha$ flux to select the zones
presenting the lowest spatial variations of the emission. These zones
were identified with the diffuse ionized gas. The selected upper
limits of the intensity gradient that we used are very similar to the
values used by \citet{2001PhDT...Zurita}: the median for our sample is
45~pc~cm$^{-6}$~arcsec$^{-1}$ and Zurita used
44~pc~cm$^{-6}$~arcsec$^{-1}$. We will discuss the results obtained
with both methods in this section.

Using both detection methods, diffuse emission has been detected in
all but 5 galaxies for an upper intensity cutoff of 65
pc~cm$^{-6}$. The median intensity of the DIG is 46 pc~cm$^{-6}$. This
value is typical of the brightest component of the DIG observed in
nearby galaxies ($3$-$5$ times larger than the averages given in
\citealt{1996AJ....112.1429H} and \citealt{2001PhDT...Zurita}).

The comparison of the flux arising from the diffuse ionized gas and
the total H$\alpha$ emission of the galaxy provides a useful
diagnostic of the escape of Lyman photons from HII regions. These
photons may ionize the gas in the disks of the galaxies or even escape
to the intergalactic medium. In both cases, it would be necessary to
revise the assumptions made when linking the measured emission from
HII regions to SFR, and when dealing with interaction between active
star formation and the interstellar and intergalactic medium.

\begin{figure} \plotone{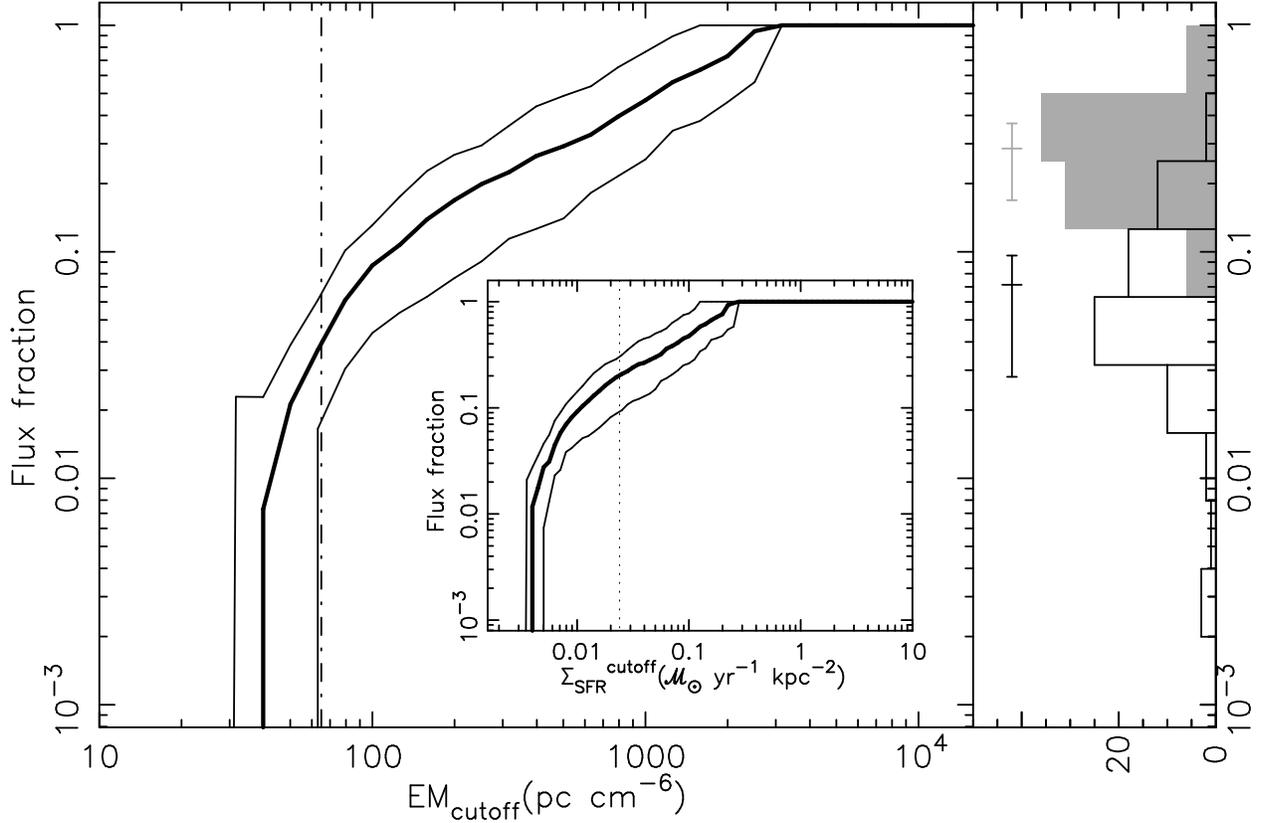}
\figcaption{\label{dig} Median contribution of the different intensity  levels
of the H$\alpha$ emission to the total flux for the galaxies in the
UCM sample (left panel). The flux fraction (vertical axis) refers to
the contribution to the total flux of the emission regions with an
intensity below a given surface brightness cutoff (horizontal axis).
The median value for the entire sample at each cutoff intensity is
shown (thick line) together with the upper and lower quartiles (thin
line). The inset plot depicts the same flux ratio against the SFR
surface density levels. For comparison, the $\Sigma_\mathrm{SFR}$
limit for the survey of $z\sim0.24$ galaxies published by
\citet{2001A&A...379..798P} is marked with a dotted line (see
text). The right panel presents the histogram of the DIG fraction of
the total flux (calculated using the pixels whose intensity is over
the global detection limit) for an EM cutoff of 65 pc~cm$^{-6}$ (empty
histogram). The gray histogram shows the same fraction but corrected
for the DIG over-imposed in the star-forming knots.} \end{figure}

\placefigure{dig} \clearpage 
\clearpage

In order to characterize the DIG in the UCM galaxies, the contribution
to the total flux of the emission zones with an intensity below a
certain surface brightness cutoff has been measured. Figure~\ref{dig}
plots this quantity against the cutoff intensity. The median values
for the 79 galaxies in our sample are shown, together with the first
and third quartiles. The upper cutoff value for the DIG (65
pc~cm$^{-6}$) is shown.

The right panel of Figure~\ref{dig} shows a histogram of the
fractional contribution of the DIG to the H$\alpha$ fluxes. Individual
values range from 0\% to 34\% (\objectname{UCM1308$+$2950}). On
average, 8\% of the total H$\alpha$ flux of a UCM galaxy comes from
pixels that were identified as DIG. Only pixels with fluxes over the
global detection limit were used in this calculation.

When using the gradient method explained above to detect the
DIG emission, we get very similar results. These
figures are consistent with previous studies on the DIG properties
\citep[e.g,][]{1996AJ....112.1429H,1999ApJ...515...97W,2001PhDT...Zurita}
and on the escape of Lyman photons from HII regions
\citep{1997MNRAS.291..827O}. 

These results do not take into account the effects of the light
scattered by the dust in the galaxies or by the optics of the
telescope. The correction for these effects has been estimated to be
5--15\% of the diffuse flux by \citet{1996AJ....112.1429H}. In
addition, the fluxes in Figure~\ref{dig} (left panel) do not contain
the contribution of the DIG located in the line of sight of the
star-forming regions, i.e., superimposed on them. We have estimated
the correction factor for this effect by substituting the intensity of
the pixels belonging to a star-forming knot by the average value of
the detected DIG pixels. This correction rises the DIG contribution to
the total flux by a factor of $\sim2$. These factors are consistent
with the ones found by
\citet{1996AJ....112.1429H}. The histogram for the flux fraction of the DIG
corrected for these effects has also been plotted in the right panel of
Figure~\ref{dig} (gray histogram).

We do not find large differences between the flux fraction of the DIG
for different morphologies. If anything, Sa galaxies show a somewhat
lower fraction ($\sim10$\%, including the corrections mentioned above)
when compared with later Hubble types (20\%).  Interacting systems and
BCDs present somewhat lower fractions. For the different spectroscopic
classes, we find that disk-{\it like} galaxies show larger fractions
($\sim20$\%) than HII-{\it like} objects (less than 10\%).

Imaging surveys of galaxies at intermediate and high redshifts are
able to detect emission-line objects to a certain flux limit which can
be expressed in units of SFR surface density
($\mathcal{M}_\sun$~yr$^{-1}$~kpc$^{-2}$). This obviously means that
objects that are fainter than that limit are not detected. But even
for objects that are detected, a significant fraction of the total
flux may be under the detection limit and thus missed.  To estimate
how important this effect may be, we have calculated the ratio of the
luminosity arising from regions with a SFR surface density below a
given cutoff value to the integrated luminosity, i.e., the cumulative
contribution of the different SFR surface density levels to the total
SFR.  Here we have calculated the total emission for these objects
adding the flux in all pixels above the detection limit.  The inset in
Figure~\ref{dig} shows the results of this calculation.  Under the
assumption that the distant galaxies under consideration are
comparable to the UCM galaxies in their star formation properties and
spatial distribution, this plot gives a correction to the detected
luminosity for each detection limit. As an example we refer to the
survey of $z\sim0.24$ H$\alpha$ emitting galaxies published by
\citet{2001A&A...379..798P}. The faintest galaxy in this survey has an
H$\alpha$ luminosity of $3.4\times10^{41}$ erg~s$^{-1}$ measured in a circular
aperture of 1.6$\arcsec$ or $\sim100$~kpc$^2$. This turns into an average SFR
surface density of $\sim0.02$~$\mathcal{M}_\sun$~yr$^{-1}$~kpc$^{-2}$, which is
shown in Figure~\ref{dig}. If we assume this value as the detection limit, this
figure reveals that the survey was missing about 20\% of the total H$\alpha$
flux in pixels with an intensity below the detection limit. Similar corrections
may be calculated for other surveys.

\subsection{H$\alpha$ luminosity function and the SFR density of the Universe}
\label{ro}

\begin{figure}
\plotone{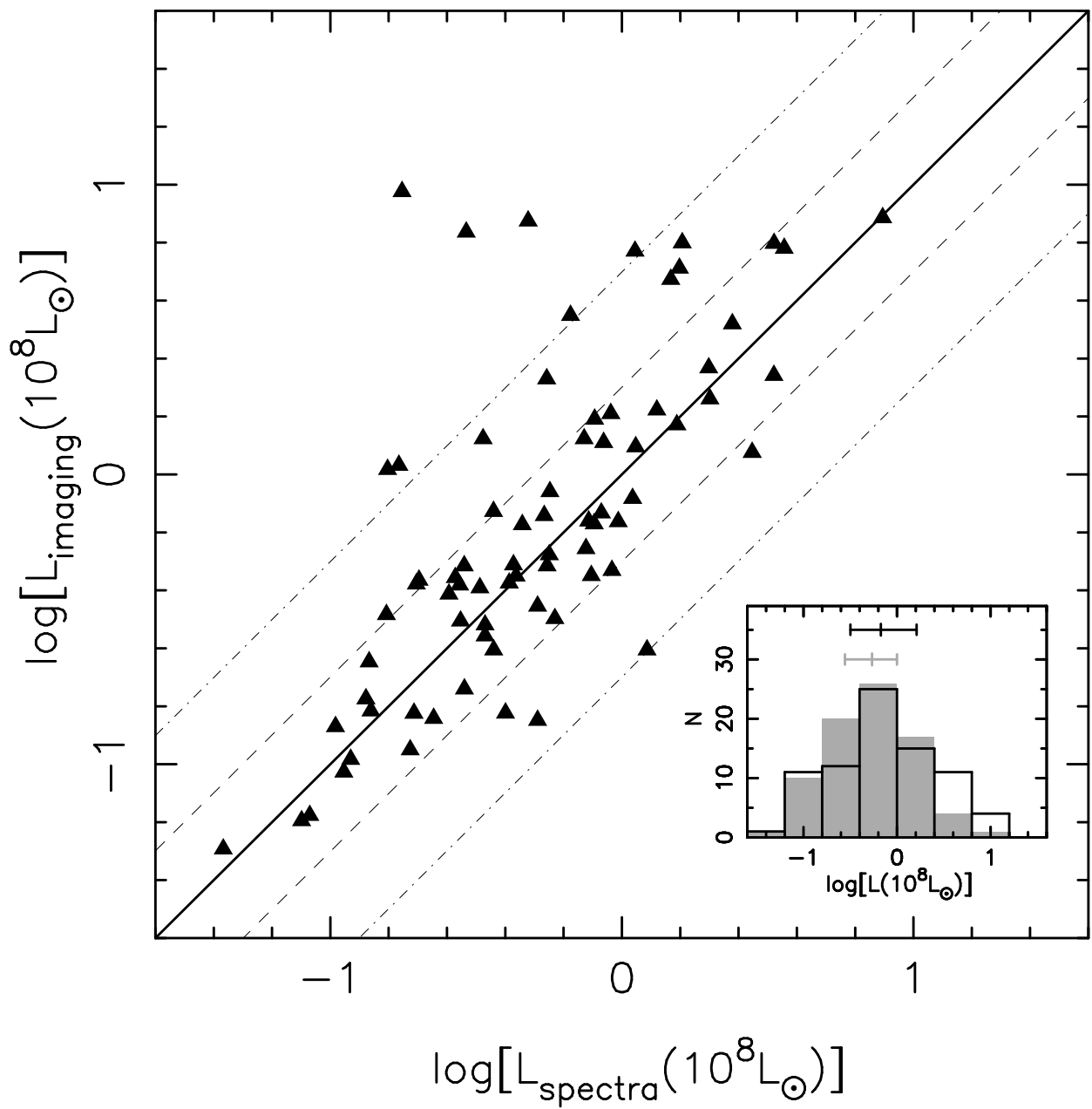}
\figcaption{\label{lum95_ima} Comparison of the H$\alpha$ luminosities 
obtained from spectroscopy (aperture corrected) and imaging. The dashed 
lines correspond to factors of  5, 2, 1/2 and 1/5.
 The inset plot shows the frequency histograms for
the imaging (empty) and spectroscopic (gray) luminosities.}
\end{figure}

The integrated luminosities measured in the H$\alpha$ images can be
used to recalculate the luminosity function and the star formation
rate density of the local Universe first published by
\citet{1995ApJ...455L...1G}. In that paper, the H$\alpha$ fluxes
measured in the spectra were converted to luminosities after an
aperture correction. This correction assumed that the $EW(\mathrm
H\alpha)$ was constant over the whole galaxy, and used integrated
Gunn-$r$ magnitudes and the spectroscopic $EW(\mathrm H\alpha)$ values
to estimate H$\alpha$ luminosities.  We compare these luminosities
with the ones measured in the H$\alpha$ images in
Figure~\ref{lum95_ima}. In the inset of that figure we also present
the histograms of the frequency distributions of the imaging and
spectroscopic luminosities (empty and gray histograms
respectively). Median values and upper/lower quartiles are indicated.

We find that, statistically, the luminosities estimated  from the spectroscopic
data  worked quite well. On average,   the aperture-corrected spectroscopic
data underestimated the total emission of the UCM Survey galaxies by
$\simeq10$\% only (the corresponding median value is 30\%).    However,
Figure~\ref{lum95_ima} shows that on an individual galaxy basis, the
spectroscopic estimates can be off by factors of up to a few.  Note that galaxy
sizes are not the determinant factor on the accuracy of the aperture
correction. The key factor is how  valid the
spectroscopically-determined  $EW(\mathrm H\alpha)$ value is for the whole
galaxy.

\placefigure{lum95_ima} \clearpage 
\clearpage

We have now recalculated the H$\alpha$ luminosity function and the SFR
density of the local Universe using the imaging-based luminosities
corrected for internal extinction. We have assumed that the galaxies
in our H$\alpha$ imaging sample are a fair subsample of the entire UCM
survey with regard to their H$\alpha$ luminosities (cf.\ section
Section~\ref{sample}), and we have followed the same techniques of the
original work of \citet{1995ApJ...455L...1G}.  For direct comparison
with the spectroscopic study, this analysis is carried out for a
cosmology with $H_{0}=50$\,km\,s$^{-1}$\,Mpc$^{-1}$, $\Omega_{\mathrm
M}=1$ and $\Lambda=0$ as well as for the cosmology used elsewhere in
this paper ($H_{0}=70$; $\Omega_{\mathrm M}=0.3$; $\Lambda=0.7$).
Using the $V/V_{max}$ method
\citep{1968ApJ...151..393S,1973ApJ...186..433H}, the Schechter
parameters of the best fit are:

\begin{minipage}[l]{7cm}
$(H_0=50,\Omega_M=1.0,\Lambda=0.0)$
\begin{eqnarray}
\alpha   & = & -1.1 \pm 0.2 \nonumber \\
\phi^*   & = & 10^{\,-3.3\pm0.2}\,{\rm Mpc}^{-3} \nonumber \\
L^*      & = &  10^{42.60\pm0.20}\,{\rm erg}\,{\rm s}^{-1} \nonumber \\ \nonumber
\end{eqnarray}
\end{minipage}
\begin{minipage}[l]{7cm}
$(H_0=70,\Omega_M=0.3,\Lambda=0.7)$
\begin{eqnarray}
\alpha   & = & -1.2 \pm 0.2 \nonumber \\
\phi^*   & = & 10^{\,-3.0\pm0.2}\,{\rm Mpc}^{-3} \nonumber \\
L^*      & = &  10^{42.43\pm0.17}\,{\rm erg}\,{\rm s}^{-1} \nonumber \\ \nonumber
\end{eqnarray}
\end{minipage}

Similar results (within the uncertainties) have been obtained using the two
more sophisticated luminosity function estimators developed by
\citet{1979ApJ...232..352S} and \citet{1988MNRAS.232..431E}. The main
difference found with the spectroscopic results is that the
characteristic H$\alpha$ luminosity $L^*$ of a local star-forming
galaxy is a factor of $\sim3$ brighter in the imaging-based study than
in the spectroscopic one. The reason for this appears to be that the
H$\alpha$ images have revealed that several large and H$\alpha$-bright
objects had their luminosities underestimated by the spectroscopic
data because their star formation was mainly located away from the
nuclear regions sampled by the slits (cf.\
figure~\ref{lum95_ima}). Similar results have been found recently by
\citet{2002MNRAS.330..621S}, based on radio observations.

The total H$\alpha$ luminosity density of the local Universe can be
calculated by integrating the luminosity function for all
luminosities. This yields values of
$10^{39.3\pm0.2}$~erg\,s$^{-1}$\,Mpc$^{-3}$ ($H_{0}=50$;
$\Omega_{\mathrm M}=1$; $\Lambda=0$) and
$10^{39.5\pm0.2}$~erg\,s$^{-1}$\,Mpc$^{-3}$ for the `concordance
cosmology'. For a fixed cosmology, the imaging-based density is
$\sim60$\% (0.2dex) larger than the spectroscopic one of
\citet{1995ApJ...455L...1G}, but note that the formal uncertainty in
the measured value is also 0.2dex.

Using the transformation from L(H$\alpha$) to SFR given in
Equation~\ref{convfac}, we obtain the SFR density of the local Universe:
$\rho_{SFR}=0.016^{+0.007}_{-0.004}$\,$\mathcal{M}_\sun$\,yr$^{-1}$\,Mpc$^{-3}$
($H_{0}=50$; $\Omega_{\mathrm M}=1$; $\Lambda=0$) and 
$\rho_{SFR}=0.029^{+0.008}_{-0.005}$\,$\mathcal{M}_\sun$\,yr$^{-1}$\,Mpc$^{-3}$
($H_{0}=70$; $\Omega_{\mathrm M}=0.3$; $\Lambda=0.7$)\footnote{  Note that the
value of the SFR density given in \citet{1995ApJ...455L...1G} 
($0.013$\,$\mathcal{M}_\sun$\,yr$^{-1}$\,Mpc$^{-3}$ for $H_{0}=50$;
$\Omega_{\mathrm M}=1$; $\Lambda=0$) was obtained  with a different
transformation factor from H$\alpha$ luminosity to SFR, so the values are not
directly comparable. The correct comparison needs to be carried out for the
H$\alpha$ luminosity densities, as we have done above.}.


\section{Summary and conclusions}

In this paper we have presented H$\alpha$ images for a
statistically-representative subsample of 79 local star-forming
galaxies. This sample has been selected from the UCM Survey galaxies
ensuring that the distributions of H$\alpha$ luminosities, equivalent
widths and spectroscopic types are representative of the entire
survey. The observations, data reduction and analysis have been
described, and we have presented results concerning galaxy properties
such as their concentration, bulge and disk emission, relationship
between newly-formed stars and underlying older population,
distribution of star-forming knots and properties of the diffuse
emission. We have also discussed the implications of our imaging
survey on the determination of the local H$\alpha$ luminosity function
and star formation rate density.  The main results of this work are:

\begin{itemize}

	\item[1.--]Typical long-slit or fiber spectroscopic
	observations of nearby galaxies such as those in the UCM
	sample ($\langle z\rangle=0.026$) can miss substantial
	fractions of the emission-line flux (typically
	50\%--70\%). Thus, to avoid applying uncertain aperture
	corrections, direct emission-line imaging (e.g., narrow-band
	H$\alpha$ imaging) can provide useful and complementary
	information to spectroscopy.

	\item[2.--]The average concentration parameter $c_{31}$
	\citep{1990A&AS...83..399G} of the UCM Survey galaxies is
	typical of disk dominated galaxies. Its distribution peaks at
	$c_{31}\sim3$. The average size of the emission regions is
	$\sim1.5$~kpc, but this size correlates with morphology in the
	sense that late-type spirals have significantly larger
	H$\alpha$-emitting regions than early types.  On average, the
	star formation event covers half of the total size of these
	galaxies.

	\item[3.--] Using a technique   akin to bulge/disk decomposition for
	the H$\alpha$ radial distribution we have robustly separated the
	contribution to the total emission into nuclear and disk components.  
	On average, a UCM galaxy harbors a star formation event with a nuclear
	component contributing  30\% of its line emission. A significant disk
	emission component is detected in the vast majority of the galaxies, 
	including objects that were spectroscopically classified as {\it
	starburst nuclei}.

	\item[4.--]In contrast with previous studies, we  do not find a
	correlation between the H$\alpha$ equivalent width and Hubble type for
	spirals.  This may be a consequence of the enhanced star-forming
	activity of the UCM galaxies (H$\alpha$-selected) with respect to
	broad-band selected ones: our sample is dominated by high-$EW(\mathrm
	H\alpha)$ galaxies.  Disk-{\it like} galaxies present considerable
	lower values of the $EW(\mathrm H\alpha)$ ($24$~\AA) than HII-{\it
	like} objects ($75$~\AA). This suggests that the underlying older
	stellar population in disk-{\it like} galaxies has a larger relative
	contribution to the galaxy luminosity than  in HII-{\it like} objects. 

	\item[5.--]A positive correlation between the SFR and the extinction
	has been found, although the correlation seems to  saturate for
	heavily-obscured objects ($\mathrm H\alpha/\mathrm H\beta\gtrsim7$
	corresponding to $E(B-V)\gtrsim0.9$).  This suggests that in  objects
	with very high extinction a large  fraction of the SFR could be
	completely hidden by dust with the Balmer line ratio only tracing those
	regions with the lowest (but still high) extinction.

	\item[6.--]The most intense star-forming knots in the UCM galaxies
	present the highest SFR surface densities and largest sizes. Moreover,
	the brightest and largest star-forming knots are  found in galaxies
	with the highest H$\alpha$ luminosities.  We have also found some
	evidence for the existence of two different star-formation regimes
	operating in `quiescent'  objects (e.g., normal spirals) and
	disturbed/interacting/irregular objects: the latter seem to have, on
	average,  star-forming regions with lower individual SFRs,  but larger 
	numbers of them so that their global SFRs can still be relatively
	high.

	\item[7.--]About 15--30\% of the total H$\alpha$ flux of a
	typical UCM galaxy comes from a diffuse ionized gas
	component. This percentage seems to be independent of the
	Hubble type.

	\item[8.--]The present H$\alpha$ imaging study has led to the
	recalculation of the H$\alpha$ luminosity function and star
	formation rate density of the local Universe. Our results
	yield an H$\alpha$ luminosity density in the local Universe of
	$10^{39.3\pm0.2}$~erg\,s$^{-1}$\,Mpc$^{-3}$ ($H_{0}=50$;
	$\Omega_{\mathrm M}=1$; $\Lambda=0$) and
	$10^{39.5\pm0.2}$~erg\,s$^{-1}$\,Mpc$^{-3}$ ($H_{0}=70$;
	$\Omega_{\mathrm M}=0.3$; $\Lambda=0.7$).  For a fixed
	cosmology, these values are 60\% larger than the densities
	calculated in \citet{1995ApJ...455L...1G}. These luminosity
	densities translate into SFR densities of
	$\rho_{SFR}=0.016^{+0.007}_{-0.004}$\,$\mathcal{M}_\sun$\,yr$^{-1}$\,Mpc$^{-3}$
	($H_{0}=50$; $\Omega_{\mathrm M}=1$; $\Lambda=0$) and
	$\rho_{SFR}=0.029^{+0.008}_{-0.005}$\,$\mathcal{M}_\sun$\,yr$^{-1}$\,Mpc$^{-3}$
	($H_{0}=70$; $\Omega_{\mathrm M}=0.3$; $\Lambda=0.7$).

\end{itemize}

\acknowledgments

PGPG wishes to acknowledge the Spanish Ministry of Education and Culture for
the reception of a {\it Formaci\'on de Profesorado Universitario} fellowship.
AAS acknowledges generous financial support from the Royal Society. AGdP
acknowledges financial support from NASA through a Long Term Space Astrophysics
grant to B.F.\ Madore. Valuable discussion with Almudena Zurita is
acknowledged. We are grateful to the anonymous referee for her/his helpful
comments and suggestions. The present work was supported by the Spanish
Programa Nacional de Astronom\'{\i}a y Astrof\'{\i}sica under grant
AYA2000-1790.

\bibliographystyle{apj}
\bibliography{referencias}

\placetable{table1}
\begin{deluxetable}{lrrccrcrrrrr}
\rotate
\tablewidth{605pt}
\tablecaption{Results of the H$\alpha$ imaging study of the UCM sample.\label{table1}}
\tablehead{\colhead{UCM name} &  \colhead{F (H$\alpha$)} & \colhead{$c_{31}$} & \colhead{$r_e$} & \colhead{$\frac{r_{80}}{r_{24.5}}$} & \colhead{EW (H$\alpha$)} & \colhead{(B/T) (H$\alpha$)} & \colhead{L (H$\alpha$)} & \colhead{$Area_T$} & \colhead{$\Sigma_{SFR}^{max}$} & \colhead{\%DIG} & \colhead{\%DIG$_c$}\\\colhead{(1)} &  \colhead{(2)} & \colhead{(3)} & \colhead{(4)} & \colhead{(5)} & \colhead{(6)} & \colhead{(7)} & \colhead{(8)} & \colhead{(9)} & \colhead{(10)} & \colhead{(11)} & \colhead{(12)}}
\startdata
0013$+$1942   &    8.6  &   2.54  &   3.1  &  0.59  &  113.4  &  0.14  &   2.57  &   70.5  &   0.498  &    4  &   22\\
0014$+$1748   &   13.6  &  10.99  &   7.8  &  0.93  &   15.6  &  0.68  &   5.68  &   71.2  &   0.470  &   32  &   48\\
0014$+$1829   &   16.6  &   6.21  &   2.3  &  0.65  &   99.6  &  0.48  &  26.26  &   99.1  &   0.075  &    4  &   24\\
0015$+$2212   &    4.2  &   3.41  &   1.0  &  0.23  &   35.2  &  0.48  &   0.54  &   29.0  &   0.037  &    6  &   20\\
0017$+$1942   &   11.2  &   1.88  &   8.9  &  0.99  &   56.1  &  0.00  &   4.93  &  113.3  &   7.250  &    4  &   23\\
0047$+$2413   &   10.7  &   5.63  &   6.9  &  0.49  &   29.3  &  0.45  &  12.66  &  132.9  &   0.464  &    7  &   23\\
0047$+$2414   &   22.5  &   3.18  &   5.5  &  0.68  &   62.0  &  0.05  &  21.27  &  241.1  &   0.368  &    4  &   19\\
0056$+$0043   &    5.3  &   3.37  &   1.5  &  0.43  &   48.5  &  0.12  &   0.86  &   26.8  &   0.357  &    2  &   18\\
0056$+$0044   &    6.6  &   9.95  &   4.2  &  0.99  &   72.2  &  0.32  &   0.57  &   69.1  &   0.214  &    9  &   39\\
0141$+$2220   &    2.8  &   4.79  &   3.0  &  1.00  &   18.6  &  0.31  &   0.64  &   26.9  &   0.213  &   12  &   33\\
0147$+$2309   &    4.0  &   2.01  &   1.9  &  0.26  &   17.5  &  0.39  &   0.95  &    5.4  &   0.195  &    0  &   62\\
0148$+$2124   &    6.5  &   5.85  &   1.2  &  0.74  &   89.7  &  0.08  &   0.57  &   32.2  &   0.153  &    1  &   28\\
0159$+$2326   &    2.5  &   2.80  &   1.1  &  0.15  &    7.5  &  0.14  &   1.19  &   15.8  &   0.150  &    6  &   15\\
1246$+$2727   &   10.0  &   3.05  &   8.1  &  0.85  &   49.5  &  0.14  &   3.42  &  179.0  &   0.149  &    7  &   38\\
1247$+$2701   &    2.5  &   1.87  &   3.8  &  0.62  &   23.5  &  0.03  &   0.40  &   36.7  &  36.985  &    7  &   36\\
1248$+$2912   &    8.4  &   5.26  &   9.1  &  0.74  &   13.6  &  0.41  &   2.06  &  118.4  &   1.686  &   10  &   30\\
1253$+$2756   &   19.9  &   1.58  &   3.7  &  0.39  &   82.0  &  0.36  &   1.22  &   65.9  &   1.061  &    1  &   14\\
1254$+$2740   &    7.9  &   3.38  &   2.0  &  0.40  &   38.8  &  0.64  &   1.26  &   63.7  &   1.001  &    2  &   25\\
1255$+$2819   &    4.0  &   1.90  &   3.3  &  0.30  &   16.9  &  0.22  &   1.71  &   54.0  &   0.875  &   10  &   31\\
1255$+$3125   &    8.0  &   4.10  &   2.0  &  0.49  &   35.7  &  0.69  &   2.64  &   42.7  &   0.722  &    2  &   25\\
1256$+$2701   &    5.2  &   4.45  &   8.8  &  0.99  &   63.1  &  0.04  &   1.16  &   79.9  &   0.710  &   11  &   49\\
1256$+$2722   &    2.1  &   3.28  &   3.5  &  0.76  &   22.9  &  0.07  &   1.65  &   29.8  &   0.443  &   13  &   41\\
1256$+$2910   &    0.7  &   2.15  &   1.2  &  0.11  &    2.9  &  1.00  &   1.85  &    8.5  &   0.432  &   11  &   33\\
1257$+$2808   &    1.7  &  10.99  &   1.1  &  0.91  &   10.0  &  0.43  &   0.58  &    7.6  &   0.431  &   12  &   39\\
1258$+$2754   &    6.8  &   2.31  &   4.2  &  0.50  &   28.2  &  0.21  &   6.21  &   90.4  &   0.376  &    7  &   26\\
1300$+$2907   &    9.0  &   3.61  &   2.4  &  0.66  &  155.1  &  0.52  &   4.12  &   16.6  &   0.317  &    0  &   39\\
1301$+$2904   &   12.8  &   2.45  &   6.4  &  0.90  &   59.8  &  0.55  &   2.59  &  195.2  &   0.216  &   15  &   45\\
1302$+$2853   &    1.8  &   1.52  &   2.5  &  0.32  &   10.4  &  0.00  &   0.55  &   25.5  &   1.125  &    5  &   11\\
1303$+$2908   &   13.0  &   1.66  &   6.9  &  0.99  &  170.6  &  0.00  &   2.02  &   78.6  &   0.805  &    3  &   16\\
1304$+$2808   &    7.0  &   2.75  &   4.7  &  0.44  &   20.7  &  0.51  &   0.70  &   83.7  &   0.746  &    8  &   28\\
1304$+$2818   &    6.4  &   2.57  &   5.3  &  0.55  &   22.9  &  0.02  &   0.95  &   93.3  &   0.434  &   17  &   58\\
1304$+$2830   &    1.1  &   2.53  &   1.3  &  0.48  &   47.9  &  0.24  &   0.20  &   12.5  &   0.130  &    2  &   33\\
1306$+$2938   &   14.6  &   2.04  &   4.2  &  0.45  &   39.1  &  0.20  &   3.16  &  118.3  &   0.122  &    4  &   23\\
1306$+$3111   &    3.2  &   2.23  &   2.7  &  0.36  &   21.8  &  0.01  &   1.59  &   59.4  &   0.104  &    8  &   23\\
1308$+$2950   &    9.1  &  10.99  &  13.9  &  0.99  &    9.4  &  0.55  &  19.71  &   41.9  &   0.063  &   34  &   62\\
1308$+$2958   &    9.3  &   2.34  &  10.3  &  0.83  &   19.0  &  0.12  &   5.08  &  118.8  &   1.559  &   24  &   63\\
1314$+$2827   &    3.9  &   1.88  &   1.9  &  0.17  &   19.5  &  0.48  &   1.87  &   32.8  &   1.239  &    4  &   15\\
1324$+$2651   &   18.1  &   3.93  &   1.6  &  0.33  &   41.0  &  0.89  &   8.93  &   81.7  &   0.613  &    3  &   17\\
1324$+$2926   &    6.4  &   2.46  &   2.0  &  0.54  &  202.6  &  0.17  &   0.43  &   34.5  &   0.487  &    0  &   15\\
1331$+$2900   &    5.3  &   3.60  &   1.0  &  0.72  &  919.4  &  0.01  &   1.56  &   22.5  &   0.341  &    0  &   13\\
1431$+$2854   &    8.5  &   2.47  &   5.3  &  0.61  &   31.3  &  0.26  &  28.67  &  127.5  &   0.250  &    6  &   44\\
1431$+$2947   &    3.3  &   3.10  &   1.9  &  0.58  &  106.3  &  0.56  &   0.36  &   34.2  &   0.201  &    5  &   26\\
1432$+$2645   &    5.9  &   7.48  &   8.9  &  0.45  &   12.6  &  0.38  &   4.77  &   56.3  &   0.160  &   23  &   41\\
1440$+$2521N  &    2.7  &   4.91  &   1.2  &  0.33  &   20.7  &  0.69  &   2.82  &   33.6  &   0.153  &   17  &   23\\
1440$+$2521S  &    5.0  &   7.45  &   4.0  &  0.99  &   45.7  &  0.61  &   1.85  &   49.6  &   0.129  &    4  &   49\\
1443$+$2548   &   15.1  &   3.04  &   5.4  &  0.72  &   59.8  &  0.04  &  18.04  &  201.9  &   0.129  &    2  &   27\\
1444$+$2923   &    4.1  &   6.22  &   7.1  &  0.99  &   42.1  &  0.85  &   1.60  &   33.6  &   0.124  &   17  &   73\\
1513$+$2012   &   24.0  &   3.71  &   2.7  &  0.42  &   80.6  &  0.51  &  24.00  &   99.3  &   0.106  &    0  &   15\\
1537$+$2506N  &   17.8  &   3.96  &   1.1  &  0.15  &   29.8  &  0.80  &   4.56  &   60.0  &   0.095  &    4  &   15\\
1537$+$2506S  &   13.8  &   3.10  &   2.5  &  0.38  &   63.7  &  0.39  &   2.63  &   65.0  &   0.082  &    3  &   15\\
1612$+$1308   &    8.2  &   2.98  &   1.0  &  0.46  &  414.5  &  0.00  &   0.24  &   18.4  &   0.073  &    0  &   13\\
1646$+$2725   &    3.3  &   2.92  &   2.2  &  0.99  &  144.8  &  0.04  &   1.69  &   35.5  &   0.071  &    5  &   25\\
1647$+$2727   &    7.3  &   2.70  &   2.5  &  0.71  &   84.5  &  0.36  &   8.19  &   56.8  &   0.064  &    7  &   30\\
1647$+$2729   &   14.7  &   2.51  &   4.5  &  0.72  &   59.2  &  0.04  &  22.57  &  163.2  &   0.060  &    3  &   25\\
1647$+$2950   &    8.1  &   5.61  &   3.3  &  0.44  &   17.4  &  0.29  &   6.98  &  110.8  &   0.925  &    6  &   28\\
1648$+$2855   &   25.4  &   1.53  &   3.5  &  0.37  &   78.5  &  0.22  &   8.41  &  123.5  &   0.696  &    2  &   10\\
1656$+$2744   &    3.6  &   2.80  &   0.8  &  0.20  &   38.0  &  0.75  &   2.76  &   18.7  &   0.511  &   10  &   13\\
1657$+$2901   &    4.5  &   2.14  &   3.5  &  0.81  &   53.0  &  0.14  &   2.85  &   56.5  &   0.459  &    6  &   26\\
2238$+$2308   &   25.8  &   2.36  &   4.8  &  0.36  &   31.1  &  0.13  &  24.12  &  134.4  &   0.444  &    6  &   32\\
2249$+$2149   &    4.3  &   8.07  &   8.1  &  0.80  &   11.5  &  0.57  &  36.23  &   40.8  &   0.396  &   16  &   57\\
2250$+$2427   &   16.2  &   4.47  &   1.4  &  0.35  &   45.2  &  0.54  &  29.43  &  110.0  &   0.338  &    6  &    7\\
2251$+$2352   &    7.0  &   2.14  &   2.2  &  0.40  &   48.9  &  0.10  &   1.34  &   49.1  &   0.325  &    4  &   14\\
2253$+$2219   &    9.4  &   2.72  &   2.7  &  0.61  &   44.5  &  0.37  &   3.34  &   76.8  &   0.294  &    2  &   22\\
2255$+$1654   &    2.6  &   4.07  &   5.4  &  0.83  &   12.5  &  0.33  &   2.12  &   40.9  &   0.272  &   22  &   50\\
2255$+$1926   &    2.4  &   5.07  &   2.3  &  0.91  &   26.8  &  0.12  &   0.25  &   19.7  &   0.261  &    4  &   43\\
2255$+$1930N  &   13.7  &   2.49  &   2.9  &  0.28  &   34.1  &  0.22  &   5.08  &   68.5  &   0.223  &    3  &   12\\
2255$+$1930S  &    8.1  &   2.72  &   2.5  &  0.49  &   39.3  &  0.19  &   1.48  &   74.8  &   0.191  &    5  &   21\\
2258$+$1920   &   10.4  &   2.68  &   4.7  &  0.58  &   35.6  &  0.05  &   1.79  &  102.0  &   0.178  &    3  &   16\\
2304$+$1640   &    3.6  &   2.55  &   1.8  &  0.52  &  101.9  &  0.00  &   0.52  &   30.8  &   0.175  &    3  &   26\\
2307$+$1947   &    3.9  &   9.41  &   1.4  &  0.79  &   21.5  &  0.39  &   1.06  &   38.0  &   0.170  &   16  &   34\\
2313$+$2517   &   11.9  &   7.70  &   3.8  &  0.44  &   11.1  &  0.44  &  13.87  &   94.0  &   0.164  &    0  &   27\\
2315$+$1923   &    5.3  &   3.50  &   0.9  &  0.30  &   96.8  &  0.76  &   5.94  &   24.0  &   0.159  &    4  &   12\\
2316$+$2457   &   35.5  &   7.34  &   7.6  &  0.56  &   30.8  &  0.38  &  23.09  &  290.3  &   0.156  &    6  &   17\\
2316$+$2459   &    6.7  &   4.74  &   6.6  &  0.54  &   21.2  &  0.34  &  13.55  &   90.7  &   0.141  &   15  &   37\\
2325$+$2318   &  150.9  &   2.32  &   8.9  &  0.64  &   72.5  &  0.00  &  11.39  &  619.3  &   0.140  &    0  &    9\\
2326$+$2435   &   13.6  &   4.65  &   3.8  &  0.71  &  100.4  &  0.27  &   1.71  &   86.5  &   0.138  &    5  &   17\\
2327$+$2515N  &    8.8  &   4.90  &   1.8  &  0.48  &   49.6  &  0.00  &   1.61  &   25.6  &   0.133  &    0  &   29\\
2327$+$2515S  &   21.0  &   2.90  &   1.7  &  0.22  &   86.2  &  0.00  &   6.39  &   35.1  &   0.132  &    0  &   14\\
2329$+$2427   &    2.0  &   9.93  &   1.4  &  0.53  &    3.8  &  0.66  &   3.99  &   15.1  &   0.121  &   14  &   42\\
\enddata
\tablecomments{Results of the H$\alpha$ imaging study. Columns stand for: (1) UCM name. (2) H$\alpha$ flux corrected for \ion{N}{2} contamination in units of $10^{-14}$~erg~s$^{-1}$~cm$^{-2}$. Median error is 16\%. (3) Concentration index $c_{31}$.  Median error is 8\%. (4) Effective radius in arcsec. Median error is 7\%. (5) Ratio between the H$\alpha$ emitting region (radius of the circular region containing 80\% of the total H$\alpha$ emission) and the entire galaxy sizes (radius of the $24.5$\,mag~arcsec$^{-2}$ in the R$_C$ band). Median error is 17\%. (6) Equivalent width of the H$\alpha$ image in \AA. Median error is 17\%. (7) Bulge-to-total luminosity ratio of the H$\alpha$ emission. Median error is 13\%. (8) H$\alpha$ luminosity corrected for internal extinction. Units are $10^{41}$ erg~s$^{-1}$. Median error is 16\%. (9) Area of the emitting region (square arcsec). (10) Maximum star formation rate density observed in $\mathcal{M}_\sun$~yr$^{-1}$~kpc$^{-2}$. Median error is 16\%. (11) Percentage of the DIG emission to the total H$\alpha$ flux. Median error is 14\%. (12) The same as (11) but corrected for the DIG overimposed in the star-forming knots.}
\end{deluxetable}

\end{document}